\documentclass[11pt]{article}
\usepackage{amsfonts}

\usepackage{amssymb}
\usepackage{latexsym}
\usepackage{graphicx}
\usepackage[english]{babel}
\begin{document}
\input epsf


\begin{flushright}
\end{flushright}
\vspace{20mm}
 \begin{center}
{\LARGE The quantum structure of black holes\footnote{Review submitted to Classical and Quantum Gravity. }}
\\
\vspace{18mm}
{\bf   Samir D. Mathur
\\}
\vspace{8mm}
Department of Physics,\\ The Ohio State University,\\ Columbus,
OH 43210, USA\\ 
\vspace{4mm}
 {mathur@mps.ohio-state.edu}\\
\vspace{4mm}
\end{center}
\vspace{10mm}
\thispagestyle{empty}
\begin{abstract}

We give an elementary review of black holes in string theory. We discuss black hole entropy
from string microstates and 
Hawking radiation from these states. We then review the structure of 2-charge microstates, and explore
how `fractionation' can lead to quantum effects over macroscopic length scales of order the horizon radius.

\end{abstract}
\newpage
\renewcommand{\theequation}{\arabic{section}.\arabic{equation}}

\def\p{\partial}
\def\h{{1\over 2}}
\def\be{\begin{equation}}
\def\bea{\begin{eqnarray}}
\def\ee{\end{equation}}
\def\eea{\end{eqnarray}}
\def\r{\rightarrow}
\def\tildr{\tilde}
\def\n{\nonumber}
\def\nn{\nonumber \\}
\def\t{\tilde}

\section{Introduction}
\label{intr}\setcounter{equation}{0}
 
 Classical general relativity predicts that we can make black holes. If we have enough mass in a given region of space then we have a process of `gravitational collapse'; the matter is squeezed towards an infinite curvature singularity at $r=0$ and the
  spacetime outside $r=0$ is described by the vacuum solution  (we are assuming  3+1 dimensions, no charge for the black hole)
\be
ds^2=-(1-{2M\over r})dt^2+(1-{2M\over r})^{-1} dr^2+r^2(d\theta^2+\sin^2\theta d\phi^2)
\label{one}
\ee
We use units such that  $G=1, c=1, \hbar=1$.
The surface $r=2M$ is the `horizon'; classically nothing can emerge from inside this region to the outside. 

A curious paradox emerges when we add in the ideas of quantum mechanics to this picture. Quantum field theory tells us that the vacuum has fluctuations, which may be characterized by pairs of particles and anti-particles that are being continuously  created and annihilated.  Hawking \cite{hawking} showed that when such a pair is created near the horizon of a black hole then one member of the pair can fall into the hole (reducing its mass) while the other member escapes to infinity, giving `Hawking radiation'. Eventually the black hole disappears, and we are left with this radiation. While energy has been conserved in the process, we have a problem. The radiation quanta were created from vacuum fluctuations near the horizon, and at this location there is no information in the geometry about what kind of matter made up the mass $M$ of the black hole. So the radiation quanta do not carry information about the initial matter which collapsed to make the hole. If $|\psi\rangle_i$ is the state of the initial matter which underwent gravitational collapse, then we cannot describe the final configuration of radiation by a state arising from a normal quantum mechanical evolution
\be
 |\psi\rangle_f = e^{-iHt}|\psi\rangle_i
\ee
since otherwise we {\it could} reconstruct the details of the initial matter by inverting the unitary evolution operator:
\be
|\psi\rangle_i = e^{iHt}|\psi\rangle_f
\ee
In fact Hawking found that the outgoing and infalling members of the pair are in an entangled state, and when the black hole vanishes (together with the members of the pairs that fell into the hole) then the radiation quanta left outside are entangled with `nothing'; i.e. they must be described not a pure state but by a density matrix. 

Closely associated to this problem is the `entropy puzzle'. Take a box of gas having an entropy $S$ and throw it into the black hole. Have we decreased the entropy of the Universe and violated the second law of thermodynamics? The work of Bekenstein \cite{bek} and Hawking \cite{hawking} shows that if we associate an entropy to the black hole equal to
\be
S_{Bek}={A\over 4G}
\label{twonew}
\ee
then the second law is saved; the decrease in entropy of the matter in the Universe is made up by the increase in the entropy of the black hole. But if we take (\ref{twonew}) seriously as the entropy of the hole then statistical mechanics tells us that there should be $e^{S_{Bek}}$ states of the black hole for the same mass. Can we see these states? The metric (\ref{one}) seems to be the unique one describing the endpoint of gravitational collapse; no small deformations are allowed, a fact encoded in the colloquial statement `Black holes have no hair'. The fact that we cannot find the $e^{S_{Bek}}$ states of the hole is the `entropy puzzle'. To see why this is tied to the information puzzle consider burning a piece of coal. The coal disappears and radiation is left, but there is no `information loss'. The state of the coal can be seen by examining the piece of the coal; a different piece of coal will have a different internal arrangement of atoms even though it might look similar at a coarse-grained level. The radiation leaves from the surface of the coal, so it can `see' the details of the internal structure of the coal. By contrast in the black hole the radiation leaves from the horizon, a region which is locally the {\it vacuum}. The initial matter went to $r=0$, which is separated by a macroscopic distance -- the horizon radius-- from the place where the radiation is created. 

If the above Hawking process were valid then we must make a big change in our ideas of quantum theory, replacing unitary evolution of pure states by a more general theory where the generic configuration is a density matrix. Not surprisingly, considerable effort was put into looking for a flaw in the Hawking computation. But the computation proved to be remarkably robust in its basic outline, and  the `Black hole information paradox' resisted all attempts at  resolution for some thirty years. 

One may imagine that since we are using general relativity and quantum theory in the same problem, we must inevitably be led to the details of  `quantum gravity', which is a poorly understood subject. So perhaps there are many things that were not done correctly in the Hawking derivation of radiation, and there might be no paradox. But we cannot escape the problem so easily. The radiation is derived from the behavior of vacuum fluctuations at the horizon, where the geometry (\ref{one}) is completely smooth; the curvature length scale  here is $\sim M$, the radius of the black hole, and can be made arbitrarily large by considering holes of larger and larger $M$. When does quantum gravity become relevant? With $G,c,\hbar$ we can make a unit of length -- the `planck length 
\be
l_p=[{G\hbar\over c^3}]^\h\sim 10^{-33} ~ cm
\ee
(we have assumed 3+1 dimensions and put in the usual values of the fundamental constants.) When the curvature length scale becomes of order $l_p$ then the  concept of spacetime as a smooth manifold must surely break down in some way. 
But for a solar mass black hole the curvature length scale at the horizon is $\sim 3~km$ and for black holes like those at the centers of galaxies it is  $\sim 10^8 ~km$, the same order as the curvature here on earth. It thus appears that we need not know anything about quantum gravity, and the simple rules of `quantum fields on curved space' would be sufficient to see how vacuum fluctuations at the horizon evolve to become radiation. It is these rules that Hawking used, and these lead to the information paradox. 

Despite the above argument suggesting that quantum gravity is not relevant, there have been numerous attempts to find fault with the Hawking computation. Many of these attempts try to use the fact that in the Schwarzschild coordinate system used in (\ref{one}) there is an infinite redshift between the horizon and infinity, so low energy quanta at infinity appear very energetic to a local observer at the horizon (see for example \cite{thooft,ps}). But this infinite redshift just signals a breakdown of the Schwarzschild coordinate system at the horizon, and a different set of coordinates -- the Kruskal coordinates-- cover both the exterior and the interior of the horizon and show no pathology at the horizon. So we need to understand the physics behind such approaches in further detail. My personal opinion on this count is that the Hawking argument can be phrased in the following more precise way. Suppose that we assume

 (a)\quad All quantum gravity effects are confined to within a fixed length scale like the planck length or string length.

 (b)\quad The vacuum is unique.

\noindent Then Hawking radiation will occur in a way that will lead to the violation of quantum mechanics.

The arguments for this phrasing will be given elsewhere.  But accepting for the moment this version of the Hawking argument  we can ask which of the assumptions breaks down if quantum mechanics is to be valid. We will describe several computations in  string theory that suggest that (a) is incorrect. How can this happen, if $l_p$ is the natural length scale made out of $G,c, \hbar$? If we scatter two gravitons off each other, then quantum gravity effects would indeed become important only when the wavelength of the gravitons dropped to a microscopic length like the string length or planck length. But a large black hole is made of a large number of quanta $N$. Is the length scale of quantum gravity $l_p$ or is it $N^\alpha l_p$ for some $\alpha>0$. We will argue, using some computations in string theory,  that the latter is true, and that the length scale $N^\alpha l_p$ is of order the horizon radius, a macroscopic length. This, if true, would  alter the picture of the black hole interior completely. It would also remove
the information paradox, since we will have a horizon sized `fuzzball', instead of the metric (\ref{one}) which is `empty space' near the horizon. Radiation leaving from the surface of the `fuzzball' can carry the information contained in the fuzzball just as radiation leaving from the surface of a piece of coal carries information about the state of the coal.

In this review we will go over some of the basic understanding of black holes  that has been obtained using string theory.
We will see how to make black holes in string theory, and how to understand their entropy and Hawking radiation. We will construct explicitly the interiors for 2-charge extremal holes. These will give us a `fuzzball' type interior conjectured above. 
We will then use qualitative arguments to suggest that all black holes have such a `fuzzball' description for the interior of their horizon.

\bigskip

{\bf Note:}  The goal of this review is to initiate the reader to some of the older work on black holes in string theory, and to show how it connects to some of the ideas developing now. We do not seek to actually review in any detail the current advances being made in the area. In particular we do not discuss 3-charge systems, in which considerable progress has been achieved over the past couple of years. We list some of these advances at the end of this article, with the hope that the reader will be inclined to delve further into the literature on the subject.

   \begin{figure}[htbp]
   \begin{center}
   \includegraphics[width=4in]{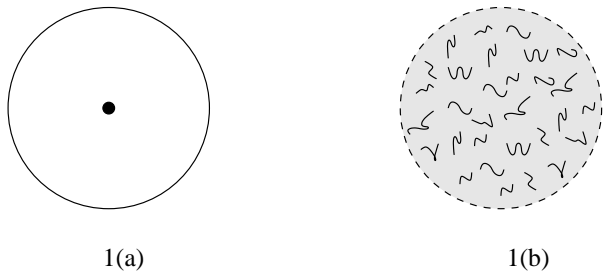}
   \caption{(a) The conventional picture of a black hole \quad (b) the proposed picture -- state information is distributed throughout the `fuzzball'. }
   \label{fig1}
   \end{center}
   \end{figure}

\section{Making black holes in string theory}
\label{maki}\setcounter{equation}{0}

For concreteness, let us start with type IIA string theory. We must make our black hole from the objects in the theory. These
objects are the following. First we have the massless graviton, present in any theory of gravity. We also have the elementary string, which we call an NS1 brane. Among the higher branes, we have an NS5 brane, and D0,D2,D4,D6,D8 branes.
We can also make Kaluza-Klein monopoles just using the gravitational field; these look like branes extending in the directions transverse to the KK-monopole.

To make a black hole we would like to put a large mass at one location. Let us start by using the elementary string, the NS1. Left to itself it will shrink and collapse to a massless quantum, so we compactify a direction to a circle $S^1$ and wrap the NS1 on this circle. To get a large mass we let the winding number be $n_1\gg 1$. It is important that we take a {\it bound} state of these $n_1$ strings, since otherwise we will end up making many small black holes instead of one big black hole. The bound state in this case is easily pictured: We let the string wrap around the circle $n_1$ times before joining back to form a closed loop; thus we have one long `multiwound' string of total length $2\pi R n_1$ where $R$ is the radius of the $S^1$.
The supergravity solution produced by such a string is
\bea
ds^2_{string}&=&H_1^{-1}[-dt^2+dy^2]+\sum_{i=1}^8 dx_idx_i\\
e^{2\phi}&=&H_1^{-1}\\
H_1&=&1+{Q_1\over r^6}
\eea
Here $ds^2_{string}$ is the 10-D string metric, $y$ is the coordinate along the $S^1$ and $x_i$ are the 8 spatial directions transverse to the string. At $r\rightarrow 0$ the dilaton $\phi$ goes to $-\infty$  and the length of the $y$ circle is seen to go to zero. The geometry does not have a horizon at any nonzero $r$, and if we say that the horizon occurs at $r=0$ then we find that the area of this horizon (measured in the Einstein metric) is zero. Thus we get $S_{Bek}=0$.

This vanishing of $S_{Bek}$ is actually consistent with the microscopic  count. The NS1 brane is in an oscillator ground state, so its only degeneracy comes from the zero modes of the string, which give 128 bosonic and 128 fermionic states. Thus we have $S_{micro}=\ln[256]$ which does not grow with   $n_1$. Thus in the macroscopic limit $n_1\rightarrow \infty$ we would write $S_{micro}=0$ to leading order, which agrees with $S_{Bek}$.

Let us go back and see why we failed to make a  black hole with nonzero area. Consider the NS1 brane as an M2 brane of M theory; this M2 brane wraps the directions $x_{11},y$. A brane has tension along its worldvolume directions, so it squeezes the cycles on which it is wrapped. Thus the length of the $x_{11}$ circle goes to zero at the brane location $r=0$, which shows up as $\phi\rightarrow -\infty$ in the IIA description. Similarly, we get a vanishing of the length of the $y$ circle in the M theory description. On the other hand if we had some directions that are compact and transverse to a brane then they would tend to expand; this happens because the flux radiated by the brane has  energy  and this energy  is lower if the flux is spread over a larger volume. 

In computing the area of the horizon we can take two equivalent approaches: 

(a) We can just look at the D noncompact directions, and find the Einstein metric (after dimensional reduction) for these noncompact directions. We compute the area $A_D$ in this metric and use the Newton constant $G_D$ for $D$ dimensions to get $S_{Bek}=A_D/4G_D$. 

(b) We can compute the area of the horizon in the full 11-D metric of M theory, and use the Newton constant for 11-D to get
$S_{Bek}=A_{11}/4 G_{11}$. In the IIA description we can compute the area of the horizon in the 10-D Einstein metric and write $S_{Bek}=A^E_{10}/4 G_{10}$. 

It is easy to check that the two computations give the same result. Let us follow (b). Then we can see that the vanishing of the $x_{11}$ and $y$ circles in the above 1-charge solution will make the 11-D horizon area vanish, and give $S_{Bek}=0$.

\subsection{Two charges}
\label{twoc}

To avoid the shrinking of the direction $x_{11}$ we can take  M5 branes and place them transverse to the direction $x_{11}$; this gives  NS5 branes in the IIA theory. To wrap the  five spatial worldvolume directions of  the NS5 branes we need more compact directions, so let us compactify a $T^4$ in addition to the $S^1$  and wrap the NS5 branes on this $T^4\times S^1$. We still have the NS1 branes along $y$, but note that with the additional compactifications the power of $r$ occurring in $H_1$ changes. We get
\bea
ds^2_{string}&=&H_1^{-1}[-dt^2+dy^2]+H_5\sum_{i=1}^4 dx_idx_i+\sum_{a=1}^4 dz_adz_a\nn
e^{2\phi}&=&{H_5\over H_1}\nn
H_1&=&1+{Q_1\over r^2}, ~~~~~\qquad H_5=1+{Q_5\over r^2}
\eea
 The $T^4$ is parametrized by $z_a, a=1\dots 4$. $Q_5$ is proportional to $n_5$, the number of NS5 branes. Note that the dilaton stabilizes to a  constant as $r\rightarrow 0$; this reflects the stabilization of the $x_{11}$ circle. Note that the $T^4$ also has a finite volume at $r=0$ since the NS5 branes cause it to shrink (their worldvolume is along the $T^4$) while the NS1 branes cause it to expand (they are transverse to the $T^4$). But the horizon area in the Einstein metric  is still zero; this can be seen in the M theory description from the fact that the NS1  (M2) and the NS5 (M5)  both wrap the $y$ circle and cause it to shrink to zero at $r=0$.

\subsection{Three charges}
\label{thre}
 
 To stabilize the $y$ circle we add {\it momentum} charge P along the $y$ circle. If we have $n_p$ units of momentum along $y$ then the energy of these modes is $n_p/R$, so their energy is {\it lower} for larger $R$. This contribution will therefore counterbalance the effect of the NS1, NS5 branes for which the energies were linearly proportional to $R$. We get
 \bea
ds^2_{string}&=&H_1^{-1}[-dt^2+dy^2+K(dt+dy)^2]+H_5\sum_{i=1}^4 dx_idx_i+\sum_{a=1}^4 dz_adz_a\n \\ 
e^{2\phi}&=&{H_5\over H_1}\n \\
H_1&=&1+{Q_1\over r^2}, ~~~~~~~~~~\qquad H_5=1+{Q_5\over r^2}, ~~~~~~~~~~~\qquad K={Q_p\over r^2}
\label{tenp}
\eea
This metric has a horizon at $r=0$. We will compute the area of this horizon in the 10-D string metric, and then convert it to the area in the Einstein metric. 

 Let us write the  metric in the noncompact directions  in polar coordinates and examine it near $r=0$
 \be
 H_5 \sum dx_idx_i= H_5(dr^2+r^2d\Omega_3^2)\approx Q_5[{dr^2\over r^2}+d\Omega_3^2]
 \ee
 Thus the area of the transverse $S^3$ becomes a constant at $r\r 0$ 
 \be
 A_{S^3}^{string} = (2\pi^2)Q_5^{3\over 2}
 \ee
 The length of the $y$ circle at  $r\r 0$ is
 \be
 L_y^{string}=(2\pi R) ({K\over H_1})^{1\over 2}= 2\pi R {Q_p^{1\over 2}\over Q_1^{1\over 2}}
 \ee
 Let the coordinate volume spanned by the $T^4$ coordinates $z_a$ be $(2\pi)^4 V$. The volume of $T^4$ at $r\rightarrow 0$ is
 \be
 V_{T^4}^{string}=(2\pi)^4 V 
 \ee
 Thus the area of the horizon at $r=0$ is
 \be
 A^{string}= A_{S^3}^{string} L_y^{string}V_{T^4}^{string}=(2\pi^2)(2\pi R)((2\pi)^4 V)Q_1^{-{1\over 2}}Q_5^{3\over 2}Q_p^{1\over 2}
 \ee
The 10-D Einstein metric $g^E_{ab}$ is related to the string metric $g^S_{ab}$ by
\be
g^E_{ab}=e^{-{\phi\over 2}}g^S_{ab}={H_1^{1\over 4}\over H_5^{1\over 4}}g^S_{ab}
\ee
At $r\rightarrow 0$ we have $e^{2\phi}={Q_5\over Q_1}$, which gives for the area of the horizon in the Einstein metric
\be
A^E=({g^E_{ab}\over g^S_{ab}})^4A^{string}={Q_1\over Q_5}A^{string}=(2\pi^2)(2\pi R)((2\pi)^4 V)(Q_1Q_5Q_p)^{1\over 2}
\label{area3charge}
\ee
The Newton constant of the 5-D noncompact space $G_5$ is related to the 10-D Newton constant $G_{10}$ by
\be
G_5={G_{10}\over (2\pi R) ((2\pi)^4 V)}
\ee
We can thus write the Bekenstein entropy as
\be
S_{Bek}={A^E\over 4G_{10}}={(2\pi^2)(2\pi R)((2\pi)^4 V)(Q_1Q_5Q_p)^{1\over 2}\over 4 G_{10}}={(2\pi^2)(Q_1Q_5Q_p)^{1\over 2}\over 4G_5}
\label{two}
\ee
We next express the $Q_i$ in terms of the integer charges
\bea
Q_1&=& { g^2 \alpha'^3\over V} n_1\nn
Q_5&=&\alpha' n_5\nn
Q_p&=&{g^2\alpha'^4\over VR^2}n_p
\label{sixt}
\eea
We have
\be
G_{10}=8\pi^6 g^2\alpha'^4
\ee
Substituting in (\ref{two}) we find
\be
S_{Bek}=2\pi(n_1n_5n_p)^{1\over 2}
\label{threeqp}
\ee
Note that  the moduli $g, V, R$ have all cancelled out. This fact is crucial to the possibility of reproducing this entropy by some microscopic calculation. In the microscopic description we will have a bound state of the charges $n_1, n_5, n_p$ and we will be counting the degeneracy of this bound state. But since we are looking at BPS states this degeneracy will not depend on the moduli. 

\subsection{Dualities}
\label{dual}

We have used three charges above: NS1 branes wrapped on $S^1$, NS5 branes wrapped on $T^4\times S^1$, and momentum P along $S^1$. If we do a T-duality in one of the directions of $T^4$ we do not change any of the charges, but reach type IIB string theory. We can now do an S-duality which gives
\be
NS1\, NS5\, P~\stackrel{\textstyle S}{\rightarrow} ~ D1\, D5\,P
\ee
Dualities can also be used to permute the three charges among themselves in all possible ways. For example four T-dualities along the four $T^4$ directions will interchange the D1 with the D5, leaving P invariant. Another set of dualities can map NS1-NS5-P to  P-NS1-NS5. Since we will make use of this map later, we give it explicitly here (the direction $y$ is called $x^5$ and the $T^4$ directions are called $x^6\dots x^9$)
\bea
&NS1\, NS5\, P\,(IIB)&~\stackrel{\textstyle T_5}{\rightarrow}~ P\, NS5 \, NS1\, (IIA)\nonumber \\
&&~\stackrel{\textstyle T_6}{\rightarrow}~ P\, NS5 \, NS1\,(IIB)\nonumber \\
&&~\stackrel{\textstyle S}{\rightarrow}~ P\, D5\, D1\,(IIB)\nonumber \\
&&\stackrel{\textstyle T_{6789}}{\rightarrow}~ P\, D1\, D5\,(IIB)\nonumber \\
&&~\stackrel{\textstyle S}{\rightarrow}~ P\, NS1 \, NS5 \,(IIB)
\label{twop}
\eea
If we keep only the first two charges in the above sequence then we see that the NS1-NS5 bound state is dual to the P-NS1 state. This duality will help us understand the geometric structure of the NS1-NS5 system, since P-NS1 is just an elementary string carrying vibrations. This duality was also profitably used in \cite{vw}.

\section{The microscopic count of states}
\label{them}\setcounter{equation}{0}

We have already seen that for the one charge case (where we had just the string NS1 wrapped on a circle $n_1$ times) we get
$S_{micro}=\ln[256]$. This  entropy does not grow with the winding number $n_1$ of the string, so from a macroscopic perspective we  get $S_{micro}\approx 0$. 

Let us now consider two charges, which we take to be NS1 and P; by the above described dualities this is equivalent to taking any two  charges from the set NS1-NS5-P or from the set D1-D5-P. The winding number of the NS1 is $n_1$ and the number of units of momentum is $n_p$. It is important that we consider the {\it bound} state of all these charges. If we break up the charges into two separate bound states then we would be describing {\it two} black holes rather than the single black hole that we wish to consider.

For  charges NS1-P it is easy to identify what states are bound. First we must join all the windings of the NS1 together to make a single string; this `long string' loops $n_1$ times around $S^1$ before joining back to its starting point. The momentum P must also be bound to this long string. If the momentum was {\it not} bound to the NS1 then it would manifest itself as massless quanta of the IIA theory (graviton, gauge fields etc) rotating around the $S^1$. When the momentum P is {\it bound} to the NS1 then it takes the form of traveling waves on the NS1. Thus the bound state of NS1-P is just a single `multiwound' string wrapped on $S^1$ with waves traveling in one direction along the $S^1$.  

But there are several ways to carry the same momentum P on the NS1. We can partition the momentum among different harmonics of vibration, and this gives rise to a large number of states for a given choice of $n_1, n_p$. 
Since we are looking at BPS states, we do not change the count of states by taking $R$ to be vary large. In this
limit we have small transverse vibrations of the NS1. We can take the DBI action for the NS1, choose the static gauge, and obtain an action for the vibrations that is just quadratic in the amplitude of vibrations. The vibrations travel at the speed of light along the direction $y$. Different Fourier modes separate and  
each Fourier mode is described by a harmonic oscillator. The total length of the NS1 is
\be
L_T=2\pi R n_1
\ee
Each excitation of the Fourier mode $k$ carries momentum and energy 
\be
p_k={2\pi k\over L_T}, ~~~e_k=|p_k|
\label{component1}
\ee
The total momentum on the string can be written as
\be
P={n_p\over R}={2\pi n_1n_p\over L_T}
\label{momentum1}
\ee
For later use we will work with the more general case where we have momentum modes traveling in both directions along the string. The extremal case was first discussed in \cite{sv} and the near extremal system  in \cite{callanmalda}. We will follow the notation of \cite{dmcompare} for later convenience. We
have a gas of excitations  with given total energy $E$ and total
momentum $P$ along  the string direction. Using a canonical ensemble, the energy $E$ is determined
 by an inverse temperature $\beta$ and the momentum $P$
by a chemical potential $\alpha$ as follows. Let there be
$m_r$ particles with energy $e_r$ and momentum  $p_r$. Define a partition function ${\cal Z}$ by
\be
\label{foone}{{\cal Z} = e^h = \sum_{states} {\rm exp}~[ -\beta\sum_r
m_r e_r - \alpha\sum_r m_r p_r]}
\ee
Then $\alpha, \beta$ are determined by requiring
\be
\label{fotwo}{ E = -{\partial h \over \partial \beta}~~~~~~~~~
P = -{\partial h \over \partial \alpha}}
\ee
The average number of particles  in the state $(e_r, p_r)$ is
then given by
\be
\label{fothree}{ \rho (e_r, p_r) = {1 \over e^{\beta e_r + \alpha p_r}
\pm 1}}
\ee
where as usual the plus sign is for fermions and the minus sign is
for bosons. Finally the entropy $S$ is given by the standard thermodynamic
relation
\be
\label{fofour}{ S = h + \alpha P + \beta E}
\ee

For the case where the gas of excitations has $f_B$ species of bosons and $f_F$
species of fermions the above quantities may be easily evaluated
\bea
P &=& {(f_B+\h f_F)L_T\pi\over 12}[{1\over (\beta + \alpha)^2}
-{1 \over (\beta - \alpha)^2}]~,~~~~
E = {(f_B+\h f_F)L_T\pi\over 12}[{1\over (\beta + \alpha)^2}
+{1 \over (\beta - \alpha)^2}]\nonumber\\
S &=& {(f_B+\h f_F)L_T\pi\over 6}[{1\over \beta + \alpha}
+{1 \over \beta - \alpha}]
\label{fofive}
\eea
Since we have massless particles in one spatial dimension, they
can be either right moving, with $e_r = p_r$ or left moving
$e_r = -p_r$. The distribution functions then become
\bea
\rho_R &=& {1 \over e^{(\beta + \alpha)e_r}\pm 1}~~~~
~~~~~~~
{\rm R}\nonumber\\
\rho_L &=& {1 \over e^{(\beta - \alpha)e_r} \pm 1}~~~~~~~~~~~
{\rm L}
\label{fosix}
\eea
Thus the combinations 
\be
T_R = {1\over(\beta + \alpha)}, ~~~ 
T_L = {1\over (\beta - \alpha)}
\ee
act as {\it effective} temperatures for the right and left moving
modes respectively. In fact all the thermodynamic quantities
can be split into a left and  a right moving piece : 
$E = E_R + E_L,~~P = P_R + P_L,~~~S = S_R + S_L$ in an obvious
notation. The various quantities $E_L, E_R, P_L, P_R, S_L, S_R$
may be read off from (\ref{fofive}). 
We get

\be
{ T_R = {\sqrt{12 E_R \over L_T\pi (f_B+\h f_F)}}
~,~~~~~~~~~~~~T_L = {\sqrt{12 E_L \over L_T\pi (f_B+\h f_F)}}}
\label{temps}
\ee
The temperature $T_H$ of the whole system (R and L movers) is given through
\be
{1\over T_H}=\beta=\h[{1\over T_R}+{1\over T_L}]
\label{thawking}
\ee
where we have used the notation $T_H$ for this temperature since we will compare it to the Hawking temperature of the black hole. The extremal state corresponds to $P_L = E_L = 0$ so that $E = P$; in this case we get $T_L=0$ and consequently $T_H=0$.

Let us now apply these results to some particular cases.

\subsection{ Extremal NS1-P}
\label{extr}

To keep contact with the black hole problem that we have set up we will continue to compactify spacetime down to 4+1 noncompact dimensions. Take the compactification $M_{9,1}\rightarrow M_{4,1}\times T^4\times S^1$. The string can vibrate in 8 transverse directions: 4 along the $T^4$ and 4 along the noncompact directions of $M_{4,1}$. Thus we have 8 bosonic degrees of freedom, and by supersymmetry, 8 fermionic partners. Thus $f_B=f_F=8$. Since we are looking at the extremal case we have all the excitations moving in the same direction, so we have only say the R movers and no L movers. This gives $E=P$, which can be achieved by letting $\alpha\r-\infty, \beta\r\infty$ with $\alpha+\beta$ finite
\be
E=P={(f_B+\h f_F)L_T\pi\over 12}[{1\over (\beta + \alpha)^2}]={2\pi n_1n_p\over L_T}
\label{entropythermal}
\ee
which gives
\be
\beta+\alpha={L_T\over \sqrt{2}\sqrt{n_1n_p}}
\ee
and
\be
S^{T^4}_{micro}=2\pi\sqrt{2}\sqrt{n_1n_p}
\label{smicro2}
\ee
For this calculation the compactification on $T^4\times S^1$ is the same as a compactification on just $S^1$, since vibrations in the $T^4$ directions are similar to those  in $R^4$. We may also consider the compactification
$M_{9,1}\rightarrow M_{4,1}\times K3\times S^1$. But IIA on K3 is dual to heterotic on $T^4$, so we can look at vibrations of the heterotic string on $T^4$. To get supersymmetric configurations we must keep the left moving supersymmetric sector in the ground state, and excite only the bosonic right movers. There are 24 transverse bosonic oscillations, so we get $f_B=24, f_F=0$. We then find
\be
S^{K3}_{micro}=4\pi\sqrt{n_1n_p}
\label{smicro2k3}
\ee

There is another equivalent language in which we can derive these microscopic  entropies. Recall that the momentum on the string is written in the form (\ref{momentum1}), where each quantum of harmonic $k$ carries the momentum (\ref{component1}).  First focus on only one of the transverse directions of vibration. If there are $m_i$ units of the Fourier harmonic $k_i$ then we need to have
\be
\sum_i m_ik_i=n_1n_p
\ee
Thus the degeneracy is given by counting  {\it partitions} of the integer $n_1n_p$.  The number of partitions of an integer $N$ is known to be
\be
P[N]~\sim~ e^{2\pi\sqrt{N\over 6}}
\label{partitions}
\ee
 We must however take into account the fact that (for the $T^4$ compactification) the momentum will be partitioned among 8 bosonic vibrations and 8 fermionic ones; the latter turn out to be equivalent to 4 bosons. Thus there are ${n_1n_p\over 12}$ units of momentum for each bosonic mode, and we must finally multiply the degeneracy in each mode. This gives for the degeneracy of states ${\cal N}$
\be
{\cal N}=[Exp(2\pi\sqrt{n_1n_p\over 72})]^{12}=Exp(2\pi\sqrt{2}\sqrt{n_1n_p})
\label{newmethod}
\ee
which again gives the entropy (\ref{smicro2}).

The 2-charge extremal entropy was first obtained in \cite{sen}, following suggestions in \cite{susskind}.

\subsection{Extremal NS1-NS5-P}\label{extr2}

Let us now ask what happens if we add in the third charge, which will be NS5 branes if the first two charges are NS1-P.
We will build a hand-waving picture for the 3-charge bound state which will be enough for our present purposes; a more systematic derivation of these properties can however be given by using an `orbifold CFT' to describe the bound state \cite{sw,lm12}.

Suppose we have only one NS5 brane. Since the NS1 brane lies along the NS5 and is bound to the NS5, we can imagine that the NS1 can vibrate inside the plane of the the NS5 but not `come out' of that plane. The momentum P will still be carried by traveling waves along the NS1, but now only four directions of vibration are allowed -- the ones inside the NS5 and transverse to the NS1. Thus $f_B$ in (\ref{entropythermal}) is 4 instead of 8. The three charge bound state is supersymmetric, so we should have 4 fermionic excitation modes as well. Then 
\be
f_B+{1\over 2} f_F=4+2=6
\ee
 But the rest of the computation can be done as for the two charge case, and we find 
 \be
 S_{micro}=2\pi\sqrt{n_1n_p}
 \ee
 Since the three charges can be permuted among each other by duality, we expect a permutation symmetric result. Since we have taken $n_5=1$ we can write
 \be
 S_{micro}=2\pi\sqrt{n_1n_5n_p}
 \ee
To understand the general case of $n_5>1$ we must get some understanding of  why the winding number $n_1$ becomes effectively $n_1n_5$ when we have $n_5$ 5-branes in the system. To do this, recall that by dualities we have the map 
\be
 NS1 (n_1) ~~ P (n_p) ~ \leftrightarrow ~ NS5 (n_1) ~~ NS1 (n_p)
 \ee
  So let us first look at NS1-P. Suppose the NS1 wraps only {\it once} around the $S^1$. The $n_p$ units of momentum are partitioned among different harmonics, with the momentum of the excitations coming in multiples of $1/R$. Now suppose the NS1 is wound $n_1>1$ times around the $S^1$. The total length of the `multiwound' string  is now $2\pi R n_1$ and the momentum now comes in multiples of 
  \be
  \Delta p=1/(n_1 R)
  \ee
   (The total momentum $n_p/R$ must still be an integer multiple of $1/R$, since this quantization must be true for any system living on the $S^1$ of radius $R$ \cite{dmfrac}.) We therefore have $n_1n_p$ units of `fractional' strength $\Delta p$ that we can partition in  different ways to get the allowed states of the system. 

Now consider the NS5-NS1 system obtained after duality. If there is only one NS5 (i.e. $n_1=1$) then we just have $n_p$ NS1 branes bound to it. Noting how different states were obtained in the NS1-P picture we expect that we can count different states  by partitioning this number $n_p$ in different ways. We can picture this by saying that the NS1 strings live in the NS5, but can be joined up to make `multiwound' strings in different ways. Thus we can have $n_p$ separate singly wound loops, or one loop wound $n_p$ times, or any other combination such that the total winding is $n_p$:
\be
\sum_i m_i k_i  = n_p
\ee
where we have $m_i$ strings with winding number $k_i$. 

 If on the other hand we have many NS5 branes ($n_1>1$) then duality indicates that the NS1 breaks up into `fractional' NS1 branes, so that there are $n_1n_p$ strands in all. These latter strands can now be grouped together in various ways so that the number of possible states is given by partitions of $n_1n_p$
 \be
 \sum_i m_i k_i=n_1n_p
 \label{six}
 \ee
In fact we should be able to reproduce the entropy (\ref{smicro2}) by counting such partitions. Let us call each `multiwound' strand in the above sum a `component string'. The only other fact that we need to know about these component strings is that they have 4 fermion zero modes coming from left movers and 4 from right movers; this can be established by a more detailed treatment of the bound states using the `orbifold CFT'. Upon quantization we get two `raising operators' and two `lowering operators' for each of the left and right sides. Starting with a ground state (annihilated by all lowering operators) 
 we can choose to apply or not apply each of the 4 possible raising operators, so we get $2^4=16$ possible ground states of the component string. Applying an even number of raising operators gives a bosonic state while applying an odd number gives a fermionic state. Each component string (with a given winding number $k$) has therefore 8 bosonic states and 8 fermionic states. 
 
 The count of possible states of the NS5-NS1 system is now just like the count for the NS1-P system. If we partition the number $n_1n_p$ as in (\ref{six}) and there are 8 bosonic and 8 fermionic states for each member in a partition, then the total number of states ${\cal N}$ will be given by (following (\ref{newmethod}))
 \be
 \ln[{\cal N}]=2\sqrt{2}\pi\sqrt{n_1n_p}
 \label{sixfollow}
 \ee
 
 With this understanding, let us return to the 3-charge system we were studying. We have $n_5$ NS5 branes and $n_1$ NS1 branes. The bound state of these two kinds of branes will generate an `effective string' which has total winding number
 $n_1n_5$ \cite{maldasuss}. This effective string can give rise to many states where the `component strings' of the state have windings $k_i$ with
 \be
 \sum m_i k_i =n_1n_5
 \label{seven}
 \ee
 
We will later use  a special subclass of states where all the component strings have the same winding $k$; we will also let each component string have the same fermion zero modes. Then the number of component strings is
\be
m={n_1n_5\over k}
\label{teight}
\ee
In the above set, one extreme case is where all component strings are singly wound
\be
k=1, ~~~m=n_1n_5
\label{eight}
\ee
The other extreme is where there is only one component string
\be
k=n_1n_5, ~~~m=1
\label{nine}
\ee

Let us now add the momentum charge $P$ to the system. We can take the NS1-NS5 bound state to be in any of the configurations (\ref{seven}), and the $n_p$ units of momentum can be distributed on the different component strings
in an arbitrary way. All the states arising in this way will be microstates of the NS1-NS5-P system, and should be counted towards the entropy. But one can see that at least for small values of $n_p$ we get a larger contribution from the case
where we have only a small number of component strings, each having a large $k_i$. To see this let $n_p=1$. First consider the extreme case (\ref{eight}). Since each component string is singly wound ($k=1$) there is no `fractionation', and we just place one unit of momentum on any one of the component strings. Further since all the component strings are alike (we chose all component strings to have the same zero modes) we do not get different states by exciting different component strings. Instead we have a state of the form
\be
|\Psi\rangle={1\over \sqrt{n_1n_5}}[({\rm component ~string~ 1 ~excited})~+~\dots ~+~({\rm component ~string~ n_1n_5 ~excited})]
\ee
The momentum mode can be in 4 bosonic states and 4 fermionic states, so we just get 8 states for the system. 

Now consider the other extreme (\ref{nine}). There is only one component string, but since it has winding $w=n_1n_5$ the one unit of momentum becomes an excitation at level $n_1n_5$ on the component string. The number of states is then given by partitioning this level into among different harmonics, and we get for the number of states
\be
{\cal N}\sim e^{2\pi\sqrt{c\over 6}\sqrt{n_1n_5}}=e^{2\pi\sqrt{n_1n_5}}
\ee
where we have used $c=6$ since we have 4 bosons and 4 fermions. This is much larger than the number of states obtained for the case $k_i=1$.

The leading order entropy for NS1-NS5-P can  be obtained by letting the NS1-NS5 be in the bound state
(\ref{nine}) and ignoring other possibilities. We put the $n_p$ units of momentum on this single long component string, getting an effective level of excitation $n_1n_5n_p$ and an entropy
\be
S_{micro}=\ln [{\cal N}] = 2\pi \sqrt{n_1n_5n_p}
\label{ten}
\ee 

We now observe that the microscopic entropy (\ref{ten}) agrees exactly with the Bekenstein entropy (\ref{threeqp})
\be
S_{micro}=S_{Bek}
\label{agree1}
\ee

This is a remarkable result, first obtained by Strominger and Vafa \cite{sv} for a slightly different system. They took the compactification $M_{4,1}\times K3\times S^1$ (i.e. the $T^4$ was replaced by K3). The case with $T^4$ was done soon thereafter by Callan and Maldacena \cite{callanmalda}.

\subsection{Non-extremal holes}\label{none}

Extremal holes offer the most rigorous connection between black hole geometries and the corresponding CFT microstates, since the energy of extremal (i.e. BPS) states depends in a known way on the charges and moduli. But since these holes have the minimum possible energy for the given charge, they do not have any `excess' energy that could be radiated away as Hawking radiation. To see this radiation we have to consider non-extremal holes, and get some microscopic picture to describe their properties.

The extremal hole has three (large) charges and no `excess' energy. We will move away from extremality in small steps, first 
keeping two large charges and some nonextremality, then one large charge and some non-extremality, and finally no large charges, a case which includes the Schwarzschild hole. While our control on the microscopics becomes less the further away we go from extremality, we will see that some general relations emerge from these studies which suggest a qualitative description for all holes.

\subsubsection{The nonextremal gravity solution}\label{then}

We continue to use the compactification $M_{9,1}\r M_{4,1}\times T^4\times S^1$. We have charges NS1, NS5, P as before, but also extra energy that gives nonextremality. The metric and dilaton are \cite{hms}
\be
ds^2_{string}=H_1^{-1}[-dt^2+dy^2+{r_0^2\over r^2}(\cosh \sigma dt+\sinh\sigma dy)^2]
+H_5[{dr^2\over (1-{r_0^2\over r^2})}+r^2d\Omega_3^2]+\sum_{a=1}^4 dz_adz_a
\label{fullmetric}
\ee
\be
e^{2\phi}={H_5\over H_1}
\ee
where
\be
H_1=1+{r_0^2\sinh^2\alpha\over r^2}, ~~~H_5=1+{r_0^2\sinh^2\gamma\over r^2}
\ee
The integer valued charges carried by this hole are
\bea
\hat n_1&=&{Vr_0^2\sinh 2\alpha\over 2g^2\alpha'^3}\label{n1}\\
\hat n_5&=&{r_0^2\sinh 2\gamma\over 2\alpha'}\label{n5}\\
\hat n_p&=&{R^2Vr_0^2\sinh 2\sigma\over 2g^2\alpha'^4}
\label{np}
\eea
The energy (i.e. the mass of the black hole) is
\be
E={RVr_0^2\over 2g^2\alpha'^4}(\cosh 2\alpha+\cosh 2\gamma+\cosh 2\sigma)
\label{energy}
\ee
The horizon is at $r=r_0$. From the area of this horizon we find the Bekenstein entropy
\be
S_{Bek}={A_{10}\over 4 G_{10}}={2\pi RV r_0^3\over g^2\alpha'^4}\cosh\alpha\cosh\gamma\cosh\sigma
\label{bek}
\ee
The Hawking temperature is 
\be
T_H=[({\p S\over \p E})_{\hat n_1, \hat n_5, \hat  n_p}]^{-1}={1\over 2\pi r_0 \cosh\alpha\cosh\gamma\cosh\sigma}
\ee

\subsubsection{The extremal limit: `Three large charges, no nonextremality'}\label{thee}

The extremal limit is obtained by taking 
\be
r_0\r 0, ~~\alpha\r\infty, ~~\gamma\r\infty, ~~\sigma\r\infty
\ee
while holding fixed
\be
r_0^2\sinh^2\alpha=Q_1, ~~r_0^2\sinh^2\gamma=Q_5, ~~r_0^2\sinh^2\sigma=Q_p
\ee
This gives the extremal hole we constructed earlier. For this case we have already checked that the microscopic entropy agrees with the Bekenstein entropy (\ref{agree1}).  It can be seen that in this limit the Hawking temperature is $T_H=0$.

\subsubsection{Two large charges $+$ nonextremality}\label{twol}

We now wish to move away from the extremal 3-charge system, towards the neutral Schwarzschild hole. For a first step, we keep two of the charges large; let these be NS1, NS5. We will have a small amount of the third charge P, and a small amount of nonextremality. The relevant limits are
\be
r_0, ~r_0e^\sigma ~\ll  r_0e^\alpha, ~r_0e^\gamma
\label{2chargelim}
\ee
Thus $\sigma$ is finite but $\alpha,\gamma\gg 1$. We are `close' to the extremal NS1-NS5 state, so we can hope that the excitations will be a small correction. The excitations will be a `dilute' gas among the large number of $\hat n_1, \hat n_5$ charges and a simple model for these excitations might give us the entropy and dynamics of the system.

The BPS mass corresponding to the $\hat n_1$ NS1 branes is
\be
M_1^{BPS}={R\hat n_1\over \alpha'}={RVr_0^2\over 2g^2\alpha'^4}\sinh 2\alpha={RVr_0^2\over 2g^2\alpha'^4}(\cosh 2\alpha-e^{-2\alpha})\approx {RVr_0^2\over 2g^2\alpha'^4}\cosh 2\alpha
\ee
The BPS mass corresponding to the $\hat n_5$ NS5 branes is
\be
M_5^{BPS}={RV\hat n_5\over g^2 \alpha'^3}={RVr_0^2\over 2g^2\alpha'^4}\sinh 2\gamma={RVr_0^2\over 2g^2\alpha'^4}(\cosh 2\gamma-e^{-2\gamma})\approx {RVr_0^2\over 2g^2\alpha'^4}\cosh 2\gamma
\label{bps5}
\ee
Thus the energy (\ref{energy}) can be written as 
\be
E= M_1^{BPS}+M_5^{BPS}+\Delta E, ~~~\Delta E\approx {RVr_0^2\over 2g^2\alpha'^4}\cosh 2\sigma
\ee
The momentum is
\be
P={\hat n_p\over R}={RVr_0^2\over 2g^2\alpha'^4}\sinh 2\sigma
\ee
Note that
\be
\Delta E+P\approx {RVr_0^2\over 2g^2\alpha'^4}e^{2\sigma}, ~~~\Delta E-P\approx {RVr_0^2\over 2g^2\alpha'^4}e^{-2\sigma}
\ee

We wish to compute the entropy (\ref{bek}) in this limit. Note that
\bea
\hat n_1&=&{Vr_0^2\over 2g^2\alpha'^3}\sinh 2\alpha\approx {Vr_0^2\over g^2\alpha'^3}\cosh^2\alpha\label{n1number}\\
\hat n_5&=&{r_0^2\over 2\alpha'}\sinh 2\gamma\approx {r_0^2\over \alpha'}\cosh^2\gamma\label{n5number}
\eea
We then find
\be
S_{Bek}\approx 2\pi\sqrt{\hat n_1\hat n_5}~[~\sqrt{{R\over 2}(\Delta E+P)}+\sqrt{{R\over 2}(\Delta E-P)}~]
\label{sbek2}
\ee

Let us now look at the microscopic description of this nonextremal state. The NS1, NS5 branes generate an `effective string' as before. In the extremal case all the excitations were right movers (R) on this effective string, so that we had the maximal possible momentum charge P for the given energy. For the non-extremal case we will have momentum modes moving in both R,L directions. Let the right movers carry $n_p$ units of momentum and the left movers $\bar n_p$ units of (oppositely directed) momentum. Then (ignoring any interaction between the R,L modes) we will have 
\be
\Delta E={1\over R}(n_p+\bar n_p), ~~~P={1\over R}(n_p-\bar n_p)
\label{npnpbar}
\ee
Since we have ignored any interactions between the R,L modes the entropy $S_{micro}$ of this `gas' of momentum modes will be the sum of the entropies of the R,L excitations. Thus using (\ref{ten}) we write
\be
S_{micro}=2\pi\sqrt{\hat n_1\hat n_5n_p}+2\pi\sqrt{\hat n_1\hat n_5\bar n_p}
\label{2chargeentropy}
\ee

But using (\ref{npnpbar}) in (\ref{sbek2}) we find 
\be
S_{micro}=2\pi\sqrt{\hat n_1\hat n_5}~[~\sqrt{{R\over 2}(\Delta E+P)}+\sqrt{{R\over 2}(\Delta E-P)}~]
\label{sbek2q}
\ee
Comparing to (\ref{sbek2}) we  find that
\be
S_{micro}\approx  S_{Bek}
\ee

We thus see that a simple model of the microscopic brane bound state describes well the entropy of this near extremal system. 

\subsubsection{One large charge $+$ nonextremality}\label{onel}

Continuing on our path to reducing the charges carried by the hole, we  now let only one charge,  NS5, be large. The relevant limit in (\ref{fullmetric}) is
\be
r_0, ~ r_0e^\alpha, ~ r_0e^\sigma ~\ll~ r_0e^\gamma
\ee
The BPS  mass for the NS5 branes is given by (\ref{bps5}), and we write
\be
E=M_5^{BPS}+\Delta E, ~~~\Delta E\approx{RVr_0^2\over 2g^2\alpha'^4}(\cosh 2\alpha+\cosh 2\sigma)
\label{extraonecharge}
\ee
Using (\ref{n5number}) in the Bekenstein entropy (\ref{bek}) gives
\be
S_{Bek}\approx {2\pi RVr_0^2\over g^2\alpha'^{7\over 2}}\sqrt{\hat n_5}\cosh\alpha\cosh\sigma
\label{onechargebek}
\ee

Let us now see how this entropy may be obtained from a microscopic model. When the NS1, NS5 charges were large then the low energy excitations were given by $P\bar P$ pairs; i.e. momentum modes and anti-momentum modes running along the effective string formed by the NS1-NS5 bound state. Now that only the NS5 charge is large, we expect that the low energy excitations will have $P\bar P$ as well as $NS1\overline{NS1}$ pairs. (Since all charges can be permuted under duality there is complete symmetry between the P and NS1 charges this time.)

Recall that when  an NS1 was bound to NS5 branes then we had postulated that the NS1 could vibrate only inside the plane of the NS5. This gives the NS1 4 transverse bosonic vibrations, and by supersymmetry 4 corresponding fermionic ones, giving a total central charge $c=4+{4\over 2}=6$. Thus the NS1 inside the NS5 branes is not a critical string. This is not a contradiction, since we do not expect it to be a fundamental string; rather it is an `effective string' whose details will have to be known to find exactly its low lying excitations. But if we assume that the string to leading order is a `free string' then we expect that excitations above the low lying ones will be given by a relation like the one for the fundamental string
\be
m^2=(2\pi R  \hat n_1 T+{\hat n_p\over R})^2+8\pi T N_L=(2\pi R  \hat n_1 T-{\hat  n_p\over R})^2+8\pi T N_R
\label{massformula}
\ee
(We have ignored the shift due to the vacuum energy since this is a small effect which depends on the details of the `effective string'; it will be subleading in what follows.) Here $\hat n_1$ is the net winding number around the $S^1$ with radius $R$, and $\hat n_p$ is the net number of units of momentum along $S^1$. 

What is the tension $T$ of this `effective string'?
When have argued that when  an NS1 brane binds to a collection of NS5 branes then the NS1 brane  becomes `fractionated' --  the NS1 breaks up into $\hat n_5$ fractional NS1 branes. Thus the tension of the `effective string'  must be
\be
T={1\over \hat n_5} T_{NS1}={1\over \hat n_5}{1\over 2\pi\alpha'}
\label{tension}
\ee
  
We are now ready to compute the microscopic entropy. For excitation levels $N_L, N_R \gg 1$ the degeneracy of string states grows like
\be
{\cal N}\sim e^{2\pi\sqrt{{c\over 6}N_R}}~e^{2\pi\sqrt{{c\over 6}N_L}}
\ee
Setting $c=6$ we find
\be
S_{micro}=\ln {\cal N}=2\pi  (\sqrt{N_L}+\sqrt{N_R})
\label{onechargemicro}
\ee
In (\ref{massformula}) the charges $\hat n_1, \hat n_p$ are given by (\ref{n1}),(\ref{np}).
We set the mass $m$ of the `effective string'  equal to the excitation energy $\Delta E$ (\ref{extraonecharge}) of the NS5 brane
\be
 m={RVr_0^2\over 2g^2\alpha'^4}(\cosh 2\alpha+\cosh 2\sigma)
 \ee
 We then find from (\ref{massformula}) and using (\ref{tension})
 \bea
 N_L&=&({RVr_0^2\over 2g^2\alpha'^{7\over 2}})^2\hat n_5\cosh^2(\alpha-\sigma)\nn
 N_R&=&({RVr_0^2\over 2g^2\alpha'^{7\over 2}})^2\hat n_5\cosh^2(\alpha+\sigma)
 \eea

Substituting in (\ref{onechargemicro}) we find
\be
S_{micro}=2\pi  (\sqrt{N_L}+\sqrt{N_R})={2\pi RVr_0^2\over g^2\alpha'^{7\over 2}}\sqrt{n_5}\cosh\alpha\cosh\sigma
\ee
Comparing to (\ref{onechargebek}) we again find \cite{malda5}
\be
S_{micro}\approx S_{Bek}
\ee

\subsubsection{No large charges}\label{nola}

This case includes in particular the Schwarzschild hole, where we set all charges to zero and have only `nonextremality'.
The system cannot be considered a near-extremal perturbation of some extremal brane system, so we do not have a simple microscopic model based on the dynamics of the corresponding branes. But we will extract some general principles from the computations done above and then extrapolate them to the general nonextremal case.

First, we have come across the idea of `fractionation': When we put momentum modes on a string of winding number $n_1$ then this momentum comes in units of ${1\over n_1}$ times  the `full' momentum unit ${1\over R}$. Similarly, NS1 branes bound to NS5 branes became `fractional'. 

Second,  we have observed that if we have large charges NS1-NS5, then nonextremal excitations are carried by (fractional) $P\bar P$ pairs. If we have only large NS5 charge then the nonextremality is carried by  a (fractional) NS1 living in the NS5 worldvolume.   This effective string could have net winding and momentum around the $S^1$, but consider for simplicity the case  $\hat n_1=\hat n_p=0$. Then the `effective string' is a `wiggling loop' on the NS5 worldvolume, with no net winding or momentum. The loop can go up and down in the $y$ direction (the direction of the $S^1$). The part which goes up can be called a part of a winding mode, and the part which comes down can be called a part of an antiwinding mode. Similarly, the `wiggles' on the string carry both positive and negative momentum along the $S^1$. So very roughly we can say that the vibrations of the  `effective string'
exhibit $P\bar P$ as well as $NS1\overline{NS1}$ pairs.

Thus in the general case of no large charges we expect that we will have $P\bar P$ pairs, $NS1\overline{NS1}$ pairs and $NS5\overline{NS5}$ pairs. Further, each  kind of brane will be `fractionated' by the other branes in the system, and the resulting degrees of freedom will give rise to the entropy of the hole.

The geometry is characterized by the integer charges $\hat n_1, \hat n_5, \hat n_p$, the energy $E$, and also three moduli: $V,R,g$ which arise from  the volume of the $T^4$, the length of the $S^1$ and the strength of the coupling. 
The entropy $S$ is a function of these 7 parameters. Let us assume that the energy of the different branes and antibranes just adds together; i.e., there is no interaction energy. There is certainly no clear basis for this assumption since we are not in a BPS situation, but we make the assumption anyway and see where it leads. If we had just NS1 branes then we can find the energy by taking the extremal limit $r_0\r 0, \alpha\r\infty$ in (\ref{n1}),(\ref{energy}) and get
\be
E_{NS1}={R\over \alpha'}n_1
\ee
In (\ref{energy}) we write the energy contributed by the NS1 branes and antibranes as the sum of the brane and antibrane contributions
\be
{RVr_0^2\over 2g^2\alpha'^4}~\cosh 2\alpha = {R\over \alpha'}(n_1+\bar n_1)
\label{sum}
\ee
The net charge is given by (\ref{n1})
\be
\hat n_1={Vr_0^2\over 2g^2\alpha'^3}~\sinh 2\alpha=n_1-\bar n_1
\label{diff}
\ee
The solution to (\ref{sum}),(\ref{diff}) is
\be
n_1={Vr_0^2\over 4g^2\alpha'^3}~e^{2\alpha}, ~~~~\bar n_1={Vr_0^2\over 4g^2\alpha'^3}~e^{-2\alpha}
\label{n1qq}
\ee
Similarly, we find
\be
n_5={r_0^2\over 4\alpha'}~e^{2\gamma}, ~~~~\bar n_5={r_0^2\over 4\alpha'}~e^{-2\gamma}
\label{n5qq}
\ee
\be
n_p={R^2Vr_0^2\over 4g^2\alpha'^4}~e^{2\sigma}, ~~~~\bar n_p={R^2Vr_0^2\over 4g^2\alpha'^4}~e^{-2\sigma}
\label{npqq}
\ee
We now observe that in terms of the $n_i, \bar n_i$ the entropy (\ref{bek}) takes a remarkably simple and suggestive form \cite{hms}
\be
S_{Bek}=2\pi(\sqrt{n_1}+\sqrt{\bar n_1})(\sqrt{n_5}+\sqrt{\bar n_5})(\sqrt{n_p}+\sqrt{\bar n_p})
\label{nonexentropy}
\ee
In the extremal limit (no antibranes) this reduces to (\ref{ten}) and in the near-extremal limit with two large charges
it reduces to (\ref{2chargeentropy}). We have seen that duality permutes the three charges, and we note that
(\ref{nonexentropy}) is invariant under this permutation. It is also interesting that  if we fix the energy $E$ (\ref{energy}) and the charges $\hat n_i$ and then require that the expression (\ref{nonexentropy}) be maximized then we get the relations (\ref{n1qq}),(\ref{n5qq}),(\ref{npqq}).

While we have no good justification for ignoring the interactions between branes and antibranes the simple form of (\ref{nonexentropy}) does suggest that for a qualitative understanding of how nonextremal holes behave we must think of the dynamics of fractional brane-antibtrane pairs. If we put energy into making a neutral black hole then we cannot think of this energy as the kinetic energy of  a `gas of gravitons'; we know that such a gas would have a much smaller entropy than $S_{Bek}$. Rather we must use the energy to create fractional brane-antibrane pairs, and we expect that the vast degeneracy of such states  will  account for the entropy. Some recent studies have also explored brane-antibrane excitations from slightly different points of view \cite{fractional}.

\subsection{The 4-charge hole}\label{the4}

So far we had compactified spacetime down to 4+1 noncompact dimensions. Let us compactify another circle, getting
$M_{9,1}\r M_{3,1}\times T^4\times S^1\times \tilde S^1$. As before we wrap  NS1 branes on $S^1$, NS5 branes on $T^4\times S^1$ and  put  momentum along $S^1$. But we can also  make KK monopoles which have  $\tilde S^1$ as the nontrivially fibred circle and which are uniform along  $T^4\times S^1$. With these charges we get a 4-charge black hole in 3+1 dimensions that is very similar to the 3-charge case in 4+1 dimensions. This time it is easier to write the Einstein metric after dimensionally reducing to the 3+1 noncompact directions since the KK-monopoles give topological twisting of the compact circle $\tilde S^1$ over the $S^2$ of the noncompact space. This metric is
\cite{4charge}
\be
ds^2_{Einstein}=-f^{-{1\over 2}} (1-{r_0\over r})dt^2+f^\h [{dr^2\over (1-{r_0\over r})}+r^2(d\theta^2+\sin^2\theta d\phi^2)]
\label{4chargemetric}
\ee
where
\be
f=(1+{r_0\sinh^2\alpha\over r})(1+{r_0\sinh^2\gamma\over r})(1+{r_0\sinh^2\sigma\over r})(1+{r_0\sinh^2\lambda\over r})
\ee
The integer valued charges carried by this hole are
\bea
\hat n_1&=&{V\tilde R r_0\sinh 2\alpha\over g^2\alpha'^3}\label{n14}\\
\hat n_5&=&{\tilde R r_0\sinh 2\gamma\over \alpha'}\label{n54}\\
\hat n_p&=&{R^2\tilde RVr_0\sinh 2\sigma\over g^2\alpha'^4}\label{np4}\\
\hat n_{KK}&=&{ r_0\sinh 2\lambda\over \tilde R}\label{nkk}
\eea
The energy (i.e. the mass of the black hole) is
\be
E={R\tilde RVr_0\over g^2\alpha'^4}(\cosh 2\alpha+\cosh 2\gamma+\cosh 2\sigma+\cosh 2\lambda)
\label{energy4}
\ee
The horizon is at $r=r_0$. From the area of this horizon we find the Bekenstein entropy
\be
S_{Bek}={A_{4}\over 4 G_{4}}={8\pi R\tilde RV r_0^2\over g^2\alpha'^4}\cosh\alpha\cosh\gamma\cosh\sigma\cosh\lambda
\label{bek4}
\ee

In the microscopic picture the NS1, NS5, KK bind together to give an `effective  string' of length 
\be
L_T=2\pi R \hat n_1\hat n_5\hat n_{KK}
\ee
The effective number of bosonic degrees of freedom is again $6$, so the near extremal dynamics is very similar to the NS1-NS5-P system.
Suppose we are studying the case with three large charges NS1, NS5, KK and a small nonextremality.
We will then add $n_p, \bar n_p$ units of momentum along the R,L directions and find a microscopic entropy, which turns out to again agree with the Bekenstein entropy found from the corresponding geometry
\be
S_{micro}=2\pi\sqrt{\hat n_1\hat n_5\hat n_{KK}n_p}+2\pi\sqrt{\hat n_1\hat n_5\hat n_{KK}\bar n_p}\approx S_{Bek}
\label{4chargeen}
\ee

\subsubsection{Summary}\label{summ2}

We have seen that simple models of extremal and near-extremal holes give good results for their entropy. A naive extrapolation to far from extremal holes works surprisingly well, and suggests that these ideas may be at least qualitatively valid for all holes. A key notion is `fractionation': Branes in a bound state `break up' other branes into fractional units, and the count of the different ways of grouping the resulting bits gives the entropy of the system. Note that if the large charges are NS1-NS5, then we get most entropy by putting the nonextremal energy into the {\it third} kind of charge; i.e., creating $P\bar P$ pairs. When we had only NS5 charge, the entropy went to creating $NS1\overline{NS1}$ and $P\bar P$ pairs. More generally, suppose we add an energy $\Delta E$ to the black hole. Playing with the ansatz (\ref{nonexentropy}) and the expression (\ref{energy}) for the energy we find that the energy $\Delta E_i$ going towards creating pairs of the $i$th kind of charge satisfies $\Delta E_i\propto 1/E_i$. Thus it is entropically advantageous to create the kind of branes that we do {\it not} have. 

\section{Absorption and emission from black holes}
\label{abso}\setcounter{equation}{0}

We have made black holes in string theory, and found that the microscopic physics of branes reproduces the Bekenstein entropy for near extremal holes. What about the dynamics of black holes? Can we compute the probability of the string state to absorb or emit quanta, and then compare this to the probability for the black hole to absorb infalling quanta or emit Hawking radiation?

\subsection{The Hawking computation}\label{theh}

Let us first look at the computation on the gravity side. Consider a spherical black hole with horizon area $A$. Suppose that we have a minimally coupled scalar in the theory
\be
\square \phi=0
\label{wave}
\ee
We wish to find the cross section for absorption of such scalars into the black hole. To do this we must solve the wave equation 
(\ref{wave}), with the following boundary conditions.  We have a plane wave incident from infinity. We put boundary conditions at the horizon which say that quanta are falling in but not coming out. Some part of the incident plane wave will be reflected from the metric around the hole and give rise to an outgoing waveform at infinity. The rest goes to the horizon, and represents the part that is absorbed. From this absorbed part we deduce the absorption cross section $\sigma$.

In general this is a hard calculation to do, but it becomes simple in the limit where the wavelength of the incident wave $\lambda$ becomes much larger than the radius of the hole. In this limit we get a universal answer \cite{unruh, dmcompare, dgm}
\be
\sigma=A
\label{area}
\ee

To compute the Hawking emission rate we must again study the wave equation in the black hole geometry. But we may just use the result  \cite{hawking,hh}  that this radiation is thermal, and thus conclude that the emission rate per unit time $\Gamma$ is related to the absorption cross section $\sigma$ by the usual thermodynamic relation
\be
\Gamma_{Hawking}=\sigma {d^dk\over (2\pi)^d}{1\over e^{\omega\over T_H}-1}
\label{radiation}
\ee
Here $d$ is the number of spatial directions in which the radiation is emitted, $T_H$ is the temperature of the hole and $\Gamma$ gives the number of quanta emitted in the given phase space range per unit time.

\subsection{The microscopic computation}\label{them2}

We would now like to see if the microscopic dynamics of branes can reproduce (\ref{radiation}). For the NS1-NS5-P hole we had the following microscopic brane picture. The NS1 branes could move inside the NS5 branes; calling the $T^4$ directions $z_1\dots z_4$ we find that the NS1 has allowed transverse vibrations in the four directions $z_1\dots z_4$. This gives 4 bosonic degrees of freedom. There are also 4 fermionic degrees of freedom, but let us ignore these for now. 

If the transverse vibrations are all traveling waves moving in the same direction along  the string then we have a BPS state with some momentum P. If we have both right and left moving waves, then we have a nonextremal state, which can decay by emitting energy. The mechanism of this decay is simple: A left moving vibration and a right moving vibration collide and leave the brane system as a massless quantum of the bulk supergravity theory. Conversely, a graviton incident on the brane bound state can convert its energy into left and right traveling waves on the string. This is of course just the same way that a guitar string emits and absorbs sound waves from the surrounding air. Let us make a simple model for the `effective string'  and see what emission rates we get. (In the following we work with branes and gravitons moving in a flat background;
the  cross section for absorption into these flat space branes  will be described in a {\it dual} way by absorption into the black hole geometry \cite{maldacena}.)

In the bulk the 10-D Einstein action  is
\be
S={1\over 2\kappa^2}\int d^{10} x \sqrt{-g} [R+\dots]
\ee
We have compactified on $T^4\times S^1$. Let us write the metric components in the $T^4$ as
\be
g_{z_iz_j}=\delta_{ij}+2\kappa \tilde h_{ij}
\ee
Then the action upto quadratic order gives
\be
S\r \int d^{10} x{1\over 2}\p_\mu \tilde h_{ij}\p^\mu \tilde h_{ij}
\ee
where the derivatives  $\p_\mu$ are nonvanishing only along in the 4+1 noncompact directions. Thus these components of the metric behave as scalars satisfying (\ref{wave}) in the noncompact space.

Now look at the brane bound state. The `effective string' stretches along the $S^1$ direction $y$. Let us model its dynamics by a Dirac-Born-Infeld type action
\be
S_{DBI}=-T\int d^2\xi \sqrt{-\det[G_{ab}]}
\ee
where 
\be
G_{ab}=g_{\mu\nu}{\p X^\mu\over \p\xi^a}{\p X^\nu\over \p\xi^b}
\label{dbi}
\ee 
is the induced metric onto the worldvolume of the effective string. We work in the static gauge
\be
X^0=t=\xi^0, ~~~X^1=y=\xi^1
\ee
In this gauge the bosonic dynamical variables are the transverse vibrations $X^i$.

Expanding (\ref{dbi}) we find 
\be
S_{DBI}\r T\int d^2\xi {1\over 2} (\delta_{ij}+ 2\kappa \tilde h_{ij})\p_aX^i\p^aX^j
\ee
where the index $a=0,1$ runs over the worldsheet coordinates $\xi^a$.
We can write $\sqrt{T}X^i\equiv \tilde X^i$ to get
\be
S_{DBI}\r \int d^2\xi {1\over 2} (\delta_{ij}+ 2\kappa \tilde h_{ij})\p_a\t X^i\p^a\t X^j
\label{inter}
\ee
Thus the total action bulk + brane is, to this order
\be
S\r \int d^{10} x{1\over 2}\p \tilde h_{ij}\p \tilde h_{ij} +  \int d^2\xi {1\over 2} (\delta_{ij}+ 2\kappa \tilde h_{ij})\p_a\t X^i\p^a\t X^j
\label{actiondbi}
\ee
Note that the tension $T$ has disappeared from these terms in the  action. This is a good thing, since it means that we don't have to worry about physical principles at this stage which will determine this tension. (For more complicated absorption processes arising from higher order terms in the action we do need the value of $T$ \cite{mathurang}.)

The effective string can only vibrate inside the $T^4$, so we have only 4 possibilities for the indices of $\tilde h_{ij}$. Let us take $\tilde h_{12}$. The kinetic term for the variable $\t h_{12}$ is $\int d^{10} x {1\over 2} \p_\mu (\sqrt{2} \tilde h_{12})\p^\mu (\sqrt{2} \tilde h_{12})$, where the fact that $\t h_{12}$, $\t h_{21}$ are the same variable gives an extra factor of $2$. Thus the interaction term in (\ref{actiondbi}) is 
\be
S_{int}=\int d^2\xi\sqrt{2}\,\kappa\, \p_a \t X^1\p^a \t X^2~ (\sqrt{2}\,\t h_{12})
\label{inter1}
\ee
When we quantize this system then we get field operators for the vibrations $\t X^1, \t X^2$ and the graviton $\t h_{12}$,
which we can expand into Fourier modes in the usual way
\bea
\hat{\t X^1}&=&\sum_{ p} {1\over \sqrt{2|p|}\sqrt{L_T}}[\hat a_{1,p} e^{i p x-i|p| t}+{\hat a_{1,p}}^\dagger e^{-i px+i|p| t}]\nn
\hat{\t X^2}&=&\sum_{ p} {1\over \sqrt{2|p|}\sqrt{L_T}}[\hat a_{2,p} e^{i p x-i|p| t}+{\hat a_{2,p}}^\dagger e^{-i p x+i|p| t}]\nn
\sqrt{2}{\hat{\t h}_{12}}&=&\sum_{\vec k} {1\over \sqrt{2\omega}\sqrt{V_9}}[\hat a_{h,\vec k} e^{i\vec k\cdot \vec x-i\omega t}+{\hat a_{h,\vec k}}^\dagger e^{-i\vec k\cdot \vec x+i\omega t}],~~~~~~(\omega=|\vec k|)
\eea
Note that the fields $\hat {\t X^1}, \hat {\t X^2}$ live on the effective string which gives a 1-dimensional box of length $L_T$. The  graviton field $ {\hat {\t h}_{12}}$ lives in the bulk. We regulate the volume of the bulk by letting the noncompact directions be in a box of volume $V_{nc}$.   The total volume of the 9-D space is then
\be
V_9=V_{nc}(2\pi R)((2\pi)^4 V)
\label{volumes}
\ee
Since the fields involved in the interaction live on different box volumes we give the steps in the following field theory computation explicitly, rather than referring the reader to a standard set of rules for computing cross sections.

The interaction (\ref{inter1}) includes the contribution
\be
S_{int}\r \sqrt{2}\,\kappa \sum_{p, \vec k}~({1\over \sqrt{2|p|}\sqrt{L_T}})({1\over \sqrt{2|p|}\sqrt{L_T}})({1\over \sqrt{2\omega}\sqrt{V_9}})L_T(2|p|^2)~\hat a_{1,p}\hat a_{2,-p}\hat a_{h,\vec k}^\dagger~e^{-2i|p|t+i\omega t}
\ee
This gives the process we seek: The vibration modes $\t X^1, \t X^2$ are annihilated and the graviton $\t h_{12}$ created.
We are looking at low energy radiation, with wavelength much larger than the compact directions, so the momentum of the outgoing graviton is purely in the noncompact directions. This sets $p^1_1=-p^1_2=p$.
The factor $(2|p|^2)$ comes from the derivatives in (\ref{inter1})
\be
p_{1a}p_2^a=-p_1^0p_2^0+p_1^1p_2^1=-2|p|^2
\ee

Focus on a particular process with a given $p>0$ and  a given $\vec k$ for the graviton. (Choosing $p>0$ implies that we have taken the right moving excitation to be $\t X^1$.) The amplitude for the decay per unit time is then
\be
R_h=\sqrt{2}\kappa |p|{1\over \sqrt{2\omega}\sqrt{V_9}}~e^{-2i|p|t+i\omega t}~=~|R_h|~e^{-2i|p|t+i\omega t}
\ee

We integrate over a large time $\Delta T$ and then take the absolute value squared to get the probability of this emission process to occur. Recall that
\be
\Big |\int_0^{\Delta T} dt\,e^{-2i|p|t+i\omega t}~\Big |^2 ~\r~2\pi \Delta T \delta (\omega-2|p|)
\ee
We must also multiply by the occupation numbers of the initial states. We then find
\be
Prob[\tilde X^1(p),\tilde X^2(-p)\rightarrow \tilde h_{12}]=|R_h|^2\rho_R(p)\rho_L(p) 2\pi \Delta T\delta (\omega-2|p|)
\ee
where $\rho_R,\rho_L$ give the occupation numbers of the right moving and left moving excitations on the effective string.

We sum over the value of $p$ in the initial  state, and approximate the sum by an integral
\be
\sum_{p}~~\rightarrow~~ {L_T\over 2\pi}\int_0^\infty dp
\ee
This gives
\bea
Prob[\tilde X^1,\tilde X^2\rightarrow \tilde h_{12}]&=&{L_T\over 2\pi}\int_0^\infty dp |R_h|^2 \rho_R(p)\rho_L(p) 2\pi \Delta T\delta (\omega-2|p|)\nn
&=&{L_T \kappa^2\omega\over 8 V_9} \rho_R({\omega\over 2})\rho_L({\omega\over 2})  \Delta T
\eea
We have
\be
\rho_R({\omega\over 2})={1\over e^{{\omega\over 2T_R}}-1}, ~~~\rho_L({\omega\over 2})={1\over e^{{\omega\over 2T_L}}-1}
\ee
where
\be
T_R=\sqrt{12 E_R\over L_T \pi (f_B+\h f_F)}=\sqrt{2 n_p\over \pi R L_T}, ~~~T_L=\sqrt{2\bar n_p\over \pi R L_T}
\ee
and we have set $E_R={ n_p\over R}, ~E_L={\bar n_p\over R}$, $f_B=f_F=4$. 
There are two additional sums to be performed. First we note that the final state graviton $\tilde h_{12}$ can be created by the interaction of a $\tilde X^1$ right mover and a $\tilde X^2$ left mover, or by the interaction of a $\tilde X^2$ right mover and a $\tilde X^1$ left mover. So we multiply the decay rate by $2$. Second, we must sum over final state gravitons. Recall that the momentum of the outgoing graviton is purely in the noncompact directions. So we get
\be
\sum_{\vec k}~~\rightarrow~~\int V_{nc}~{d^4k\over (2\pi)^4}
\ee
Putting all this together the rate of decay is \cite{dmcompare}
\bea
\Gamma_{micro}&=&\int V_{nc}~{d^4k\over (2\pi)^4}2{1\over \Delta T}~Prob[\tilde X^1,\tilde X^2\rightarrow \tilde h_{12}]\nn
&=&\int {d^4 k\over (2\pi)^4}~(2\pi\omega G_5 L_T)~\rho_R({\omega\over 2})\rho_L({\omega\over 2}) 
\label{gmicro}
\eea
where we have used $2\kappa^2=16\pi G_{10}=16\pi (2\pi R)((2\pi)^4 V) G_5$ and (\ref{volumes}).

\subsubsection{Three large charges $+$ nonextremality}\label{thre2}

First let us go to the extremal 3-charge NS1-NS5-P hole that we constructed and add a small amount of nonextremality so that the hole radiates. The area of the horizon is given to a first approximation by the area of the extremal hole, so in the Hawking emission rate (\ref{radiation}) we can use  $\sigma$ equal to the area  (\ref{area3charge}). What does our microscopic calculation give?

The 3-charge extremal hole is obtained for $n_P\ne 0$  but $\bar n_p=0$.  To go off extremality by a small amount we take  $\bar n_p$ small but nonzero.  Thus $T_L\ll T_R$. From  (\ref{thawking}) we have
\be
{1\over T_H}=\h[{1\over T_R}+{1\over T_L}]\approx\h{1\over T_L}
\ee
Consider radiation quanta with energy $\omega$ comparable to the black hole temperature
\be
\omega\sim T_L, ~~~\omega \ll T_R
\label{ineq}
\ee
Then we have
\be
\rho_R\approx {2 T_R\over \omega}
\ee
and we find (using $L_T=n_1n_5(2\pi R)$)
\bea
\Gamma&=&\int {d^4 k\over (2\pi)^4}~\rho_L(-{\omega\over 2}) 8\pi G_5\sqrt{n_1n_5n_p}\nn
&=&\int {d^4 k\over (2\pi)^4}{1\over e^{\omega\over T_H}-1}(4G_5) (2\pi\sqrt{n_1n_5n_p})
\eea
 Noting that 
\be
2\pi\sqrt{n_1n_5n_p}\approx S_{micro}={A_5\over 4G_5}
\ee
we see that
\be
\Gamma_{micro}=\int {d^4 k\over (2\pi)^4}{1\over e^{\omega\over T_H}-1} A_5
\ee
Using (\ref{area}) in (\ref{radiation}) we find \cite{dmcompare} 
\be
\Gamma_{micro}~=~\Gamma_{Hawking}
\ee

Thus we see that if we slightly excite the 3-charge extremal state then the radiation from the brane state has the same
gross behavior as the Hawking radiation from the corresponding near-extremal hole. 

\subsubsection{Two large charges $+$ nonextremality}\label{twol2}

In studying the entropy of brane states we started with the case of three large charges and worked towards lesser charges.
We follow the same approach here. Let us now take the case studied in section(\ref{twol}) where we had two large charges NS1-NS5 plus a small amount of nonextremality. Now $n_p$ is not large, rather $n_p, \bar n_p$ should be thought of as being of the same order (and much smaller than $n_1, n_5$). We have $T_H\sim T_L\sim T_R$, so we consider radiation with energy
\be
\omega\sim T_L, ~~~\omega \sim T_R
\label{ineqq}
\ee
For the Hawking result (\ref{radiation}) we   calculate the absorption cross section $\sigma$ by  solving the wave equation (\ref{wave}) in the geometry (\ref{fullmetric}) with parameters in the domain (\ref{2chargelim}), and we find \cite{maldastrom}
\be
\sigma=\pi^3 (r_0^2\sinh^2\alpha)(r_0^2\sinh^2\gamma)\omega {e^{\omega\over T_H}-1\over (e^{\omega\over 2 T_R}-1)(e^{\omega\over 2 T_L}-1)}
\ee
Using (\ref{radiation}) the Hawking emission rate will be
\be
\Gamma_{Hawking}=\int {d^4 k\over (2\pi)^4}~\pi^3 (r_0^2\sinh^2\alpha)(r_0^2\sinh^2\gamma)\omega {1\over (e^{\omega\over 2 T_R}-1)(e^{\omega\over 2 T_L}-1)}
\label{greybody}
\ee
This does not have the standard black body form as a function of temperature $T_H$, so we say that the radiation is dressed by `greybody factors'. Such greybody factors are always expected to occur when the wavelength of the radiated wave becomes comparable to some length scale like the diameter of the radiating body.

Now look at the microscopic expression (\ref{gmicro}). Substituting the values of $G_5, L_T$ we have
\be
\Gamma_{micro}=\int {d^4 k\over (2\pi)^4}~({\pi^3 n_1n_5\alpha'^4 g^2\over V})\omega {1\over (e^{\omega\over 2 T_R}-1)(e^{\omega\over 2 T_L}-1)}
\ee
Using 
\be
r_0^2\sinh^2\alpha\approx  {g^2\alpha'^3\over V}n_1, ~~~r_0^2\sinh^2\gamma\approx {\alpha'}n_5
\ee
we again find \cite{maldastrom}
\be
\Gamma_{micro}=\Gamma_{Hawking}
\ee

This is very interesting, because we have reproduced the greybody factors found  in the classical computation  (\ref{greybody}) by a string theory calculation. The classical  result  had factors that correspond in the string computation to left moving and right moving particle densities. It is as if the classical geometry `knew' that there was an effective string description of the black hole. 

\subsubsection{One large charge $+$ nonextremality}\label{onel2}

Let us now consider the case where we have one large charge (NS5) plus nonextremality. At leading order the absorption cross section for minimal scalars gives the absorption cross section $\sigma=A$ (eq. (\ref{area})) and the emission rate is  given by using  (\ref{radiation}). On the microscopic side we saw that the states are accounted for by a fractional tension string vibrating in the plane of the NS5 branes. We can compute the absorption by such a string, and we find  \cite{klebanovmathur}
\be
\sigma_{micro}=A
\ee
which again implies $\Gamma_{micro}=\Gamma_{Hawking}$. A similar result holds for the case of 3+1 noncompact dimensions when we have 2 large charges + nonextremality.

\subsubsection{Summary}

We see that the degrees of freedom that of brane bound states that gave an entropy equalling the Bekenstein entropy also give radiation that agrees with the Hawking radiation from the corresponding holes. This is important, because it indicates that we have identified the correct physical degrees of freedom in the state.
Similar computations of absorption have been done for some other brane bound states (see for example
\cite{klebanov}) and agreement with the Hawking rate obtained.

\section{Constructing the microstates}
\label{cons}\setcounter{equation}{0}

We have seen that  string theory gives us a count of microstates which agrees with the Bekenstein entropy, and using the dynamics of weakly coupled branes we also correctly reproduce Hawking radiation.  But to solve the information problem we need to know  what these microstates {\it look} like. We want to understand the structure of states in the coupling domain where we get the black hole. This is in contrast to a  count at $g=0$ which can give us the correct {\it number} of states (since BPS states do not shift under changes of $g$) but will not tell us what the inside of a black hole looks like. 

At this stage we already notice a puzzling fact. For the three charge case we found $S_{micro}=S_{Bek}=2\pi\sqrt{n_1n_5n_p}$. But suppose we keep only two of the charges, setting say $n_5=0$. Then the Bekenstein entropy $S_{Bek}$  becomes zero; this is why we  had to take three charges to get a good black hole. But the microscopic entropy for two charges NS1-P was $S_{micro}=2\pi\sqrt{2}\sqrt{n_1n_p}$, which is nonzero. 

One might say that the 2-charge case is just not a system that gives a good black hole, and should be disregarded in our investigation of black holes. But this would be strange, since on the microscopic side the entropy of the 2-charge system arose in a very similar way to that for the three charge system; in each case we partitioned among harmonics the momentum 
on a string or `effective string'. We would therefore like to take a closer look at the gravity side of the problem for the case of two charges. 

We get the metric for NS1-P by setting to zero the $Q_5$ charge in  (\ref{tenp}). With a slight change of notation we write the metric as ($u=t+y, ~v=t-y$)
 \bea
ds^2_{string}&=&H[-dudv+Kdv^2]+\sum_{i=1}^4 dx_idx_i+\sum_{a=1}^4 dz_adz_a\n \\
B_{uv}&=&{1\over 2}[H-1]\n \\
e^{2\phi}&=&H\n \\
H^{-1}&=&1+{Q_1\over r^2}, ~~\qquad K={Q_p\over r^2}
\label{naive}
\eea
We will call this metric the {\it naive} metric for NS1-P. This is because we will later argue that this metric is not produced by any configuration of NS1, P charges. It is a solution of the low energy supergravity equations away from $r=0$, but just because we can write such a solution does not mean that the singularity at $r=0$ will be an allowed one in the full string theory. 

What then are the singularities that {\it are} allowed? If we start with flat space, then string theory tells us that excitations around flat space are described by configurations of various fundamental objects of the theory; in particular, the fundamental string. We can wrap this string around  a circle like the $S^1$ in our compactification. We have also seen that we can wrap this string $n_1$ times around the $S^1$ forming a bound state. For $n_1$ large this configuration will generate the solution which has only NS1 charge
 \bea
 ds^2_{string}&=&H[-dudv]+\sum_{i=1}^4 dx_idx_i+\sum_{a=1}^4 dz_adz_a\n \\
B_{uv}&=&{1\over 2}[H-1]\n \\
e^{2\phi}&=&H\n \\
H^{-1}&=&1+{Q_1\over r^2}
\label{f1}
\eea
This solution is also singular at $r=0$, but this is a singularity that we must accept since the geometry was generated by a source that exists in the theory. One may first take the limit $g\r 0$ and get the string wrapped $n_1$ times around $S^1$ in flat space. Then we can increase $g$ to a nonzero value, noting that we can track the state under the change since it is a BPS state. If $n_1$ is large and we are not too close to $r=0$ then (\ref{f1}) will be a good description of the solution corresponding to the bound state of $n_1$ units of NS1 charge. 

Now let us ask what happens when we add P charge. We have already seen that in the bound state NS1-P the momentum P will be carried as traveling waves on the `multiwound' NS1. Here we come to the most critical point of our analysis: {\it There are no longitudinal vibration modes of the fundamental string NS1}. Thus all the momentum must be carried by transverse vibrations. But this means that the string must bend away from its central axis in order to carry the momentum, so it will not be confined to the location $r=0$ in the transverse space. We will shortly find the correct solutions for NS1-P, but we can already see that the solution (\ref{naive}) may be incorrect since it requires the NS1-P source to be at a point $r=0$ in the transverse space.

The NS1 string has many strands since it is multiwound. When carrying a generic traveling wave these strands will separate from each other. We have to find the metric created by these strands.  Consider the bosonic excitations, and for the moment restrict attention to the 4 that give bending in the noncompact directions $x_i$. The wave carried by the NS1 is then described by a transverse displacement profile $\vec F(v)$, where $v=t-y$. The metric for a single strand of the string carrying such a wave is known \cite{wave}
 \bea
ds^2_{string}&=&H[-dudv+Kdv^2+2A_i dx_i dv]+\sum_{i=1}^4 dx_idx_i+\sum_{a=1}^4 dz_adz_a\nn
B_{uv}&=&{1\over 2}[H-1], ~~\qquad B_{vi}=HA_i\nn
e^{2\phi}&=&H\nn
H^{-1}(\vec x ,y,t)&=&1+{Q_1\over |\vec x-\vec F(t-y)|^2}\nn
K(\vec x ,y,t)&=&{Q_1|\dot{\vec F}(t-y)|^2\over |\vec x-\vec F(t-y)|^2}\nn
A_i(\vec x ,y,t)&=&-{Q_1\dot F_i(t-y)\over |\vec x-\vec F(t-y)|^2}
\label{fpsingle}
\eea
Now suppose that we have many strands of the NS1 string,  carrying different vibration profiles $\vec F^{(s)}(t-y)$. 
While the vibration profiles are different, the strands all carry momentum in the same direction $y$. In this case the strands are mutually BPS and the metric of all the strands can be obtained by superposing the harmonic functions arising in the solutions for the individual strands. Thus we get 
 \bea
ds^2_{string}&=&H[-dudv+Kdv^2+2A_i dx_i dv]+\sum_{i=1}^4 dx_idx_i+\sum_{a=1}^4 dz_adz_a\n \\
B_{uv}&=&{1\over 2}[H-1], ~~\qquad B_{vi}=HA_i\n \\
e^{2\phi}&=&H\n \\
H^{-1}(\vec x, y,t)&=&1+\sum_s{Q_1^{(s)}\over |\vec x-\vec F^{(s)}(t-y)|^2}\n \\
K(\vec x, y,t)&=&\sum_s{Q_1^{(s)}|\dot{\vec F}^{(s)}(t-y)|^2\over |\vec x-\vec F^{(s)}(t-y)|^2}\n \\
A_i(\vec x,y,t)&=&-\sum_s{Q_1^{(s)}\dot F^{(s)}_i(t-y)\over |\vec x-\vec F^{(s)}(t-y)|^2}
\label{fpmultiple}
\eea

Now consider the string that we actually have in our problem. 
 We can open up the multiwound string by going to the $n_1$ fold cover of $S^1$. Then the string is described by the profile $\vec F(t-y)$, with $0\le y<2\pi R n_1$.  The part of the string in the range $0\le y<2\pi R$ gives one strand in the actual space, the part in the range $2\pi R\le y<4\pi R$ gives another strand, and so on. These different strands do not lie on top of each other in general, so we have a many strand situation as in (\ref{fpmultiple}) above. But note that the end of one strand is at the same position as the start of the next strand, so the strands are not completely independent of each other. In any case all strands are given once we give the profile function $\vec F(v)$.
 
 The above solution has a sum over strands that looks difficult to carry out in practice. But now we note that there is a simplification in the `black hole' limit which is defined by
 \be
 n_1, n_p\r \infty
 \label{fiftqq}
 \ee
 while the moduli like $g, R, V$ are held fixed. We have called this limit the black hole limit for the following reason.
 As we increase the number of quanta $n_i$ in a bound state, the system will in general change its behavior and properties. In the limit  $n_i\r\infty$ we expect that there will be a certain set of properties that will govern the system, and these are the properties that will be the universal ones that characterize  large black holes (assuming that the chosen charges do form a black hole).
 
 The total length of the NS1 multiwound string is $2\pi n_1 R$. Consider the gas of excitations considered in section(\ref{them}). The energy of the typical excitation is
 \be
e\sim T\sim {\sqrt{n_1n_p}\over L_T}
\ee
so that the generic quantum is in a harmonic 
\be
k\sim \sqrt{n_1n_p}
\label{thir}
\ee
on the multiwound NS1 string. So the wavelength of the vibration is 
 \be
 \lambda\sim {2\pi Rn_1\over \sqrt{n_1n_p}}\sim {\sqrt{n_1\over n_p}}R
 \label{fift}
 \ee
The generic state of the string will be a complicated wavefunction arising from excitations of all the Fourier modes of the string, so it will not be well described by a classical geometry. We will first take some limits to get good classical solutions, and use the results to estimate the `size' of the generic `fuzzball'. Let us take a state where the typical wavenumber is much smaller than the value (\ref{thir})
 \be
 {k\over \sqrt{n_1n_p}}\equiv \alpha \ll1
 \ee
Then the wavelength of the vibrations is much longer than the length of the compactification circle
\be
\lambda={2\pi R n_1\over k}={2\pi R \over \alpha}\sqrt{n_1\over n_p}\gg2\pi R
\ee
where we have assumed that $n_1, n_p$ are of the same order. 

When executing its vibration the string will move in the transverse space across a coordinate distance
\be
\Delta x\sim |\dot{\vec F}|\lambda
\ee
But the distance between neighboring strands of the string will be
\be
\delta x=|\dot{\vec F}|(2\pi R)
\ee
We thus see that
\be
{\delta x\over \Delta x}\sim \sqrt{n_p\over n_1}~\alpha\ll1
\ee

We can therefore look at the metric at points that are not too close to any one of the strands, but that are still in the general region occupied
by the vibrating string
\be
|\vec x-\vec F(v)|\gg\delta x
\ee
(It turns out that after we dualize to NS1-NS5 the geometry is smooth at the location of the strands; we will see this in an explicit example below, and for a generic discussion see \cite{lmm, fuzz}.)  In this case neighboring strands give very similar contributions to the harmonic functions in (\ref{fpmultiple}), and we may replace the sum by an integral
\be
\sum_{s=1}^{n_1} \r \int _{s=0}^{n_1} ds = \int_{y=0}^{2\pi R n_1}{ds\over dy} dy
\ee
Since the length of the compacification circle is $2\pi R$ we have
\be
{ds\over dy}={1\over 2\pi R}
\ee
Also, since the vibration profile is a function of $v=t-y$ we can replace the integral over $y$ by an integral over $v$. Thus we have
\be
\sum_{s=1}^{n_1}\r {1\over 2\pi R}\int_{v=0}^{L_T }dv
\ee
where 
\be
L_T=2\pi R n_1
\label{fseven}
\ee
is the total range of the $y$ coordinate on the multiwound string. Finally, note that
\be
Q_1^{(i)}={Q_1\over n_1}
\ee
We can then write the NS1-P solution as
 \bea
ds^2_{string}&=&H[-dudv+Kdv^2+2A_i dx_i dv]+\sum_{i=1}^4 dx_idx_i+\sum_{a=1}^4 dz_adz_a\nn
B_{uv}&=&{1\over 2}[H-1], ~~\qquad B_{vi}=HA_i\nn
e^{2\phi}&=&H
\label{ttsix}
\eea
where
\bea
H^{-1}&=&1+{Q_1\over L_T}\int_0^{L_T}\! {dv\over |\vec x-\vec F(v)|^2}\\
K&=&{Q_1\over
L_T}\int_0^{L_T}\! {dv (\dot
F(v))^2\over |\vec x-\vec F(v)|^2}\\
A_i&=&-{Q_1\over L_T}\int_0^{L_T}\! {dv\dot F_i(v)\over |\vec x-\vec F(v)|^2}
\label{functionsq}
\eea

\subsection{Obtaining the NS1-NS5 geometries}\label{obta}

From (\ref{twop}) we see  that we can perform S,T dualities to map the above NS1-P solutions to NS1-NS5 solutions.
For a detailed presentation of the steps (for a specific $\vec F(v)$) see \cite{lm3}. The computations are straightforward, except for one step where we need to perform an electric-magnetic duality. Recall that under T-duality a Ramond-Ramond gauge field form $C^{(p)} $
can change to a higher form $C^{(p+1)} $ or to a lower form $C^{(p-1)} $. We may therefore find ourselves with $C^{(2)}$ and $C^{(6)}$ at the same time in the solution. The former gives $F^{(3)}$ while the latter gives $F^{(7)}$. We should convert the $F^{(7)}$ to $F^{(3)}$ by taking the dual, so that the solution is completely described using only $C^{(2)}$.  Finding $F^{(3)}$ is straightforward, but it takes some inspection to find a $C^{(2)}$ which will give this $F^{(3)}$. 

 Note that we have chosen to write the classical solutions in a way where $\phi$ goes to zero at infinity, so that the true dilaton $\hat\phi$ is given by
\be
e^{\hat\phi}=ge^\phi
\ee
The dualities change the values of the moduli describing the solution. Recall that the $T^4$ directions are $x^6, x^7, x^8, x^9$, while the $S^1$ direction is $y\equiv x^5$. We keep track of (i) the coupling $g$ (ii) the value of the scale $Q_1$ which occurred in the harmonic function for the NS1-P geometry (iii) the radius $R$ of the $x^5$ circle (iv) the radius $R_6$ of the $x^6$ circle, and (v) the volume $(2\pi)^4 V$ of $T^4$.  We can start with NS1-P and reach NS5-NS1, which gives
(here we set $\alpha'=1$ for compactness)
\be\label{DualParam}
\left(\begin{array}{c}
g\\Q_1\\R\\R_6\\V
\end{array}\right)
\stackrel{\textstyle S}{\rightarrow}
\left(\begin{array}{c}
1/g\\Q_1/{g}\\R/\sqrt{g}\\R_6/\sqrt{g}\\V/g^2
\end{array}\right)
\stackrel{\textstyle T6789}{\rightarrow}
\left(\begin{array}{c}
g/V\\Q_1/{g}\\R/\sqrt{g}\\\sqrt{g}/R_6\\g^2/V
\end{array}\right)
\stackrel{\textstyle S}{\rightarrow}
\left(\begin{array}{c}
V/g\\Q_1{V}/g^2\\R\sqrt{V}/g\\\sqrt{V}/R_6\\V
\end{array}\right)
\stackrel{\textstyle T56}{\rightarrow}
\left(\begin{array}{c}
R_6/R\\Q_1{V}/g^2\\g/(R\sqrt{V})\\R_6/\sqrt{V}\\R_6^2
\end{array}\right)
\equiv
\left(\begin{array}{c}
g'\\Q_5'\\R'\\R'_6\\V'
\end{array}\right)
\label{eightt}
\ee
where at the last step we have noted that the $Q_1$ charge in NS1-P becomes the NS5 charge $Q'_5$ in NS5-NS1. 
We will also choose coordinates at each stage so that the metric goes to $\eta_{AB}$ at infinity. Since we are writing the string metric, this convention is not affected by T-dualities, but when we perform an S-duality we need to re-scale the coordinates to keep the metric $\eta_{AB}$.  In the NS1-P solution the harmonic function generated by the NS1 branes is (for large $r$) 
\be
H^{-1}\approx 1+{Q_1\over r^2}
\label{ninet}
\ee
After we reach the NS1-NS5 system by dualities the corresponding harmonic function will behave as
\be
H^{-1}\approx 1+{Q'_5\over r^2}
\label{fsix}
\ee
where from (\ref{eightt}) we see that
\be
Q'_5=\mu^2 Q_1
\ee
with
\be
\mu^2={V\over g^2}
\label{fone}
\ee
Note that $Q_1, Q'_5$ have units of $(length)^2$. Thus all lengths get scaled by a factor $\mu$ after the dualities. Note that
\be
Q'_5=\mu^2Q_1=\mu^2{g^2n_1\over V}=n_1\equiv n'_5
\label{feight}
\ee
which is the correct parameter to appear in the harmonic function (\ref{fsix}) created by the NS5 branes.

With all this, for NS5-NS1 we get the solutions \cite{lm4}
\be
ds^2_{string}={1\over 1+K}[-(dt-A_i dx^i)^2+(dy+B_i dx^i)^2]+{1\over
H}dx_idx_i+dz_adz_a
\label{qsix}
\ee
where the harmonic functions are
\bea
H^{-1}&=&1+{\mu^2Q_1\over \mu L_T}\int_0^{\mu L_T} {dv\over |\vec x-\mu\vec F(v)|^2}\nn
K&=&{\mu^2Q_1\over
\mu L_T}\int_0^{\mu L_T} {dv (\mu^2\dot
 F(v))^2\over |\vec x-\mu\vec F(v)|^2},\nonumber\\
A_i&=&-{\mu^2Q_1\over \mu L_T}\int_0^{\mu L_T} {dv~\mu\dot F_i(v)\over |\vec x-\mu\vec F(v)|^2}
\label{functionsqq}
\eea
Here $B_i$ is given by
\be
dB=-*_4dA
\label{vone}
\ee
and $*_4$ is the duality operation in the 4-d transverse  space
$x_1\dots
x_4$ using the flat metric $dx_idx_i$.

By contrast the `naive' geometry which one would write for NS1-NS5 is
\be
ds^2_{naive}={1\over (1+{Q'_1\over r^2})}[-dt^2+dy^2]+(1+{Q'_5\over
r^2})dx_idx_i+dz_adz_a
\label{d1d5naive}
\ee

\subsection{A special example}\label{aspe}

The above general solution looks rather complicated. To get a feeling for the nature of these NS1-NS5 solutions let us start by examining in detail a simple case. Start with the NS1-P solution which has the following vibration profile for the NS1 string
\be
F_1=\hat a\cos\omega v,\quad F_2=\hat a\sin\omega v, \quad F_3=F_4=0
\label{yyonePrime}
\ee
where $\hat a$ is a constant.
This makes the NS1 swing in a uniform helix in the $x_1-x_2$ plane. Choose
\be
\omega=\frac{1}{n_1R}
\ee
This makes the NS1 have just one turn of the helix in the covering space. Thus all the energy has been put in the lowest harmonic on the string.

We then find
\be
H^{-1}=1+{Q_1\over 2\pi}\int_0^{2\pi} {d\xi\over
(x_1-\hat a\cos\xi)^2+(x_2-\hat a\sin\xi)^2+x_3^2+x_4^2}
\ee
To compute the integral we introduce polar coordinates in the $\vec x $ space
\bea\label{EpolarMap}
x_1&=&{\tilde r} \sin{\tilde \theta} \cos{\tildr\phi}, ~~~\qquad x_2={\tilde r}
\sin{\tilde\theta} \sin{\tildr\phi},\nonumber \\
x_3&=&{\tilde r} \cos{\tilde\theta} \cos{\tildr\psi}, ~~\qquad x_4={\tilde r}
\cos{\tilde\theta} \sin{\tildr\psi}
\label{etwo}
\eea
Then we find
\be
H^{-1}=1+{Q_1\over
\sqrt{(\tilde r^2+\hat a^2)^2-4 \hat a^2\tilde r^2\sin^2\tilde\theta}}
\ee

The above expression simplifies if we
change from $\tilde r, \tilde\theta$ to coordinates $r,\theta$:
\bea
{\tilde r}&=& \sqrt{r^2+\hat a^2\sin^2\theta}, ~~\qquad \cos{\tilde\theta}
={r\cos\theta\over \sqrt{r^2+\hat a^2\sin^2\theta}}
\label{ethree}
\eea
(${\tildr\phi}$ and ${\tildr\psi}$ remain unchanged). Then we get
\be
H^{-1}=1+{Q_1\over r^2+\hat a^2\cos^2\theta}
\ee
Similarly we get
\be
K={\hat a^2\over n_1^2 R^2}~{Q_1\over (r^2+\hat a^2\cos^2\theta)}
\ee
With a little algebra we also find
\bea
A_{x_1}&=&{Q_1\hat a\over 2\pi R n_1}\int_0^{2\pi}{d\xi \sin\xi\over (x_1-\hat a\cos\xi)^2+(x_2-\hat a\sin\xi)^2+x_3^2+x_4^2}\nn
&=&{Q_1\hat a\over 2\pi R n_1}\int_0^{2\pi}{d\xi \sin\xi\over (\tilde r^2+\hat a^2-2\tilde r \hat a\sin\tilde\theta\cos(\xi-\tilde\phi))}
\nn
&=&{Q_1\hat a^2\over R n_1}\sin\tilde\phi {\sin\theta\over (r^2+a^2\cos^2\theta)}{1\over \sqrt{r^2+a^2}}
\eea
\bea
A_{x_2}&=&-{Q_1\hat a^2\over R n_1}\cos\tilde\phi {\sin\theta\over (r^2+a^2\cos^2\theta)}{1\over \sqrt{r^2+a^2}}
\eea
\be
A_{x_3}=0, \qquad A_{x_4}=0
\ee
We can write this in polar coordinates
\bea
A_{\tilde\phi}&=&A_{x_1}{\p x_1\over \p\tilde\phi}+A_{x_2}{\p x_2\over \p\tilde\phi}\nn
&=&-{Q_1\hat a^2\over R n_1}{\sin^2\theta\over (r^2+a^2\cos^2\theta)}
\eea
We can now substitute these functions in (\ref{ttsix}) to get the solution for the NS1-P system for the choice of profile (\ref{yyonePrime}).

Let us now get the corresponding NS1-NS5 solution. Recall that all lengths scale up by a factor $\mu$ given through (\ref{fone}).
The transverse displacement profile $\vec F$  has units of length, and so scales up by the factor $\mu$. We define
\be
a\equiv \mu \hat a 
\ee
so that
\be
\mu F_1= a\cos\omega v,\quad \mu F_2= a\sin\omega v, \quad F_3=F_4=0
\label{yyonePrimep}
\ee
Let
\be
f=r^2+a^2\cos^2\theta
\ee
The NS1 charge becomes the NS5 charge after dualities, and corresponding harmonic function becomes
\be
H'^{-1}=1+{Q'_5\over f}
\ee
The harmonic function for momentum P was
\be
K={Q_1\hat a^2\over n_1^2R^2}{1\over (r^2+\hat a^2 \cos^2\theta)}\equiv {Q_p\over (r^2+\hat a^2 \cos^2\theta)}
\ee
After dualities $K$ will change to the harmonic function generated by NS1 branes. Performing the change of scale (\ref{fone})
we find
\be
K'=\mu^2 {Q_p\over f}\equiv{Q'_1\over f}
\ee
Using the value of $Q_1$ from (\ref{sixt}) we observe that
\be
a={\sqrt{Q'_1Q'_5}\over R'}
\label{ftwo}
\ee
where $R'$ is the radius of the $y$ circle after dualities (given in (\ref{eightt})). 

To finish writing the NS1-NS5 solution we also need the functions $B_i$ defined through (\ref{vone}). In the coordinates
$r, \theta, \tilde\phi\equiv\phi, \tilde\psi\equiv\psi$ we have
\be
A_\phi=-{a\sqrt{Q'_1Q'_5}}{\sin^2\theta\over f}
\ee
We can check that the dual form is
\be
B_\psi=-{a\sqrt{Q'_1Q'_5}}{\cos^2\theta\over f}
\ee
To check this, note that the flat 4-D metric in our coordinates is
\be
dx_idx_i={f\over r^2+a^2}dr^2+fd\theta^2+(r^2+a^2)\sin^2\theta d\phi^2+r^2\cos^2\theta d\psi^2
\ee
We also have
\be
\epsilon_{r\theta\phi\psi}=\sqrt{g}=f r\sin\theta\cos\theta
\ee
We then find
\be
F_{r\psi}=\partial_r B_\psi={a\sqrt{Q'_1Q'_5}}{2r\cos^2\theta\over f^2}=-\epsilon_{r\psi\theta\phi}g^{\theta\theta}g^{\phi\phi}[\partial_\theta A_\phi]=-(*dA)_{r\psi}
\ee
\be
F_{\theta\psi}=\partial_\theta B_\psi={a\sqrt{Q'_1Q'_5}}{r^2\sin(2\theta)\over f^2}=-\epsilon_{\theta\psi r\phi}g^{rr}g^{\phi\phi}[\partial_r A_\phi]=-(*dA)_{\theta\psi}
\ee
verifying (\ref{vone}).

Putting all this in (\ref{qsix}) we find the NS1-NS5 (string) metric for the profile (\ref{yyonePrime})
\bea\label{MaldToCompare}
d{s}^2&=&-H_1^{-1}(d{t}^2-d{ y}^2)+
H_5f\left(d\theta^2+\frac{d{r}^2}{{r}^2+a^2}\right)
-\frac{2a\sqrt{Q'_1 Q'_5}}{H_1f}\left(\cos^2\theta d{ y}d\psi+
\sin^2\theta d{ t}d\phi\right)\nonumber\\
&+&H_5\left[
\left({ r}^2+\frac{a^2Q'_1Q'_5\cos^2\theta}{H_1H_5f^2}\right)
\cos^2\theta d\psi^2+
\left({ r}^2+a^2-\frac{a^2Q'_1Q'_5\sin^2\theta}{H_1H_5f^2}\right)
\sin^2\theta d\phi^2\right]\nonumber \\
&+&~dz_adz_a
\eea
where
\be\label{defFHProp}
f={r}^2+a^2\cos^2\theta,\qquad
H_1=1+{Q'_1\over f}, ~~H_5=1+{Q'_5\over f}
\ee

At large $r$ this metric goes over to flat space. Let us consider the opposite limit  $r\ll
(Q'_1Q'_5)^{1/4}$ (we write $r'=r/a$):
\bea
ds^2&=&-({r'}^2+1)\frac{a^2dt^2}{Q'_1}+{r'}^2
\frac{a^2dy^2}{Q'_1}+
Q'_5\frac{d{r'}^2}{{r'}^2+1}\nonumber\\
&+&Q'_5\left[d\theta^2+\cos^2\theta \left(d{\psi}-
\frac{ady}{\sqrt{Q'_1Q'_5}}\right)^2+
\sin^2\theta \left(d{\phi}-\frac{adt}{\sqrt{Q'_1Q'_5}}\right)^2\right]\nn
&+&dz_adz_a
\label{fthree}
\eea
Let us transform to new angular coordinates
\be
\psi'=\psi-{a\over \sqrt{Q'_1Q'_5}}y, ~~\qquad \phi'=\phi-{a\over \sqrt{Q'_1Q'_5}}t
\ee
Since $\psi,y$ are both periodic coordinates, it is not immediately obvious that the first of these changes makes sense.
The identifications on these coordinates are
\be
(\psi\r \psi+2\pi, ~~y\r y), ~~\qquad (\psi\r\psi, ~~y\r y+2\pi R')
\ee
But note that we have the relation (\ref{ftwo}), which implies that the identifications on the new variables are
\be
(\psi'\r \psi'+2\pi, ~~y\r y), ~~\qquad (\psi'\r \psi'-{a2\pi R'\over \sqrt{Q'_1Q'_5}}=\psi'-2\pi, ~~y\r y+2\pi R')
\ee
so that we do have a consistent lattice of identifications on $\psi',y$. 
The metric (\ref{fthree}) now becomes
\bea
\label{esix}
ds^2&=&Q'_5\left[
-({r'}^2+1)\frac{dt^2}{R^2}+{r'}^2
\frac{dy^2}{R^2}+
\frac{d{r'}^2}{{r'}^2+1}\right]\nonumber\\
&+&Q'_5\left[d\theta^2+\cos^2\theta d{\psi'}^2+
\sin^2\theta d{\phi'}^2\right]+dz_adz_a
\eea
This is just $AdS_3\times S^3\times T^4$. Thus the full geometry is flat at infinity, has a `throat' type region at smaller $r$
where it approximates the naive geometry (\ref{d1d5naive}), and then instead of a singularity at $r=0$ it ends in a smooth `cap'. This particular geometry, corresponding to the profile (\ref{yyonePrime}), was derived earlier in \cite{bal,mm} by taking limits of general rotating black hole solutions found in \cite{cy}. We have now obtained it by starting with the particular NS1-P profile (\ref{yyonePrime}), and thus we note that it is only one member of the complete family parametrized by $\vec F$. It can be shown \cite{lmm, fuzz},  that all the metrics of this family have the same qualitative structure as the particular metric that we studied; in particular they have no horizons, and they end in smooth `caps' near $r=0$. We depict the  2-charge NS1-NS5 microstate  geometries in Figure 2.

\subsection{`Size' of the 2-charge bound state}\label{size}

The most important point that we have seen in the above discussion is that in the NS1-P bound state the NS1 undergoes transverse vibrations that cause its strands to spread out over a nonzero range in the transverse $\vec x$ space. Thus the bound state is not `pointlike'. Exactly how big {\it is} the bound state?

We have obtained good classical solutions by looking at solutions where the wavelength of vibrations $\lambda$ was much longer than the wavelength for the generic solution. To get an estimate of the size of the generic state we will now take our classical solutions and extrapolate  to the domain where $\lambda$ takes its generic value (\ref{fift}). 

The wavelength of vibrations  for the generic state is
\be
\lambda={L_T\over k}\sim {2\pi R n_1\over \sqrt{n_1n_p}}\sim R\sqrt{n_1\over n_p}
\ee
We wish to ask how much the transverse coordinate $\vec x$ changes in the process of oscillation. Thus we set $\Delta y=\lambda$, and find
\be
\Delta x\sim |\dot{\vec F}|\Delta y\sim |\dot{\vec F}|R\sqrt{n_1\over n_p}
\ee
Note that
\be
Q_p \sim Q_1 |\dot{\vec F}|^2
\ee
which gives
\be
\Delta x\sim \sqrt{Q_p\over Q_1}R\sqrt{n_1\over n_p}\sim \sqrt{\alpha'}
\label{ffone}
\ee
where we have used (\ref{sixt}). 

For 
\be
|\vec x|\gg\Delta x
\ee
we have
\be
{1\over |\vec x-\vec F|^2}\approx {1\over |\vec x|^2}
\ee
and the solution becomes approximately the naive geometry (\ref{naive}).

We see that the metric settles down to the naive metric outside a certain ball shaped region $|\vec x|>\sqrt{\alpha'}$. 
Let us now ask an important question: What is the surface area of this ball?

First we compute the area in the 10-D string metric. Note that the metric will settle down to the naive form (\ref{naive}) quite rapidly as we go outside the region occupied by the vibrating string.  The mean wavenumber is $k\sim \sqrt{n_1n_p}$, so there are $\sim \sqrt{n_1n_p}$ oscillations of the string. There is in general no correlation between the directions of oscillation in each wavelength, so the string makes $\sim \sqrt{n_1n_p}$ randomly oriented traverses in the ball that we are investigating. This causes a strong cancellation of the leading moments like dipole, quadrupole ...etc. The surviving moments will be of very high order, an order that will increase with $n_1, n_p$ and which is thus infinite in the classical limit of large charges. 

We must therefore compute the area, in the naive metric (\ref{naive}), of the location $|\vec x| =\sqrt{\alpha'}$. Introduce polar coordinates on the 4-D transverse space
\be
d\vec x\cdot d\vec x=dr^2+r^2 d\Omega_3^2
\ee
At the location $r=\sqrt{\alpha'}$ we get from the angular $S^3$ an area
\be
A_{S^3}\sim \alpha'^{3\over 2}
\label{tsix}
\ee
From the $T^4$ we get an area
\be
A_{T^4}\sim V
\ee
From the $S^1$ we get a length
\be
L_y\sim \sqrt{HK} R \sim \sqrt{Q_p\over Q_1} R
\ee
Thus the area of the 8-D surface bounding the region occupied by the string is given, in the string metric, by
\be
A^{S}\sim \sqrt{Q_p\over Q_1}RV{\alpha'}^{3\over 2}
\ee
The area in Einstein metric will be
\be
A^E=A^{S}e^{-2\phi}
\ee
Note that the dilaton $\phi$ becomes very negative at the surface of interest
\be
e^{-2\phi}=H^{-1}\approx {Q_1\over r^2}\sim {Q_1\over \alpha'}
\label{tfive}
\ee
We thus find
\be
A^E\sim  \sqrt{Q_1Q_p}RV\alpha'^{1\over 2}\sim {g^2\alpha'^4}\sqrt{n_1n_p}
\ee
where we have used (\ref{sixt}).
Now we observe that
\be
{A^E\over 4G_{10}}\sim \sqrt{n_1n_p}\sim S_{micro}
\label{fftwo}
\ee
This is very interesting, since it shows that the surface area of our `fuzzball' region satisfies a Bekenstein type relation
\cite{lm5}.

     \begin{figure}[htbp]
   \begin{center}
   \includegraphics[width=6in]{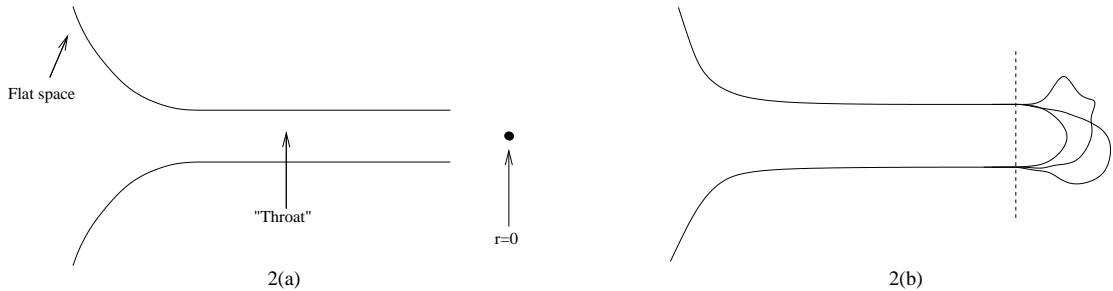}
   \caption{(a) The naive geometry of extremal NS1-NS5 \quad (b) the actual geometries; the area of the surface denoted by the dashed line reproduces the microscopic entropy.}
   \label{fig2}
   \end{center}
   \end{figure}

\subsection{Nontriviality of the `size'}

One of the key ideas we are pursuing is the following. To make a big black hole we will need to put together many elementary quanta. What is the size of the resulting bound state? One possibility is that this size is always of order planck length $l_p$ or string length $l_s$. In this case we will not be able to avoid the traditional picture of the black hole. Since the horizon radius can be made arbitrarily large, the neighborhood of the horizon will be `empty space' and the matter making the black hole will sit in a small neighborhood of the singularity. But a second possibility is that the size of a bound state {\it increases} with the number of quanta in the bound state
\be
{\cal R}\sim N^\alpha l_p
\label{ttwo}
\ee
where ${\cal R}$ is the radius of the bound state, $N$ is some count of the number of quanta in the state, and the power $\alpha$ depends on what quanta are being bound together. It would be very interesting if in every case we find that
\be
{\cal R}\sim R_H
\label{tthree}
\ee
where $R_H$ is the radius of the horizon that we would find for the classical geometry which has the mass and charge carried by these $N$ quanta. For in that case we would find that we do not get the traditional black hole; rather we get a `fuzzball' which has a size of order the horizon size. Since we do not  get a traditional horizon we do not have  the usual computation of Hawking radiation which gives information loss. The different configurations of the fuzzball will correspond to the $e^{S_{Bek}}$ states expected from the Bekenstein entropy.

For the 1-charge system we saw that the Bekenstein entropy was $S_{Bek}=0$. We also find no nontrivial size for the bound state, so the size remains order $l_p$ or $l_s$, with the exact scale depending perhaps on the choice of probe. This is consistent with the picture we are seeking, but not a nontrivial illustration of the conjecture. But the situation was much more interesting when we moved to the 2-charge case. The microscopic entropy was $S_{micro}=2\sqrt{2}\sqrt{n_1n_p}$. The size of the bound state was such that the area of the boundary satisfied a Bekenstein type relation. We had verified this relation using the 10-D metric, but we can also write it in terms of quantities in the dimensionally reduced 5-D theory
\be
{A_5\over 4 G_5}\sim \sqrt{n_1n_p}
\label{tone}
\ee
We define the 5-D planck length by
\be
l_p^{(5)}\equiv G_5^{1\over 3}
\ee
We also define the radius of the horizon from the area
\be
{\cal R}=[{A_5\over 2\pi ^2}]^{1\over 3}
\ee
The result (\ref{tone}) then translates to
\be
{\cal R}\sim (n_1n_p)^{1\over 6} l^{(5)}_p
\label{tfour}
\ee
Thus for the 2-charge system we find a manifestation of the conjectured relations (\ref{ttwo}), (\ref{tthree}).

While we see from (\ref{tfour}) that the fuzzball size ${\cal R}$ is much larger then planck length $l_p$, we have not yet compared
${\cal R}$ to the string length $l_s$. From (\ref{tsix}) we see that 
\be
{\cal R}\sim \sqrt{\alpha'}\sim l_s
\label{tseven}
\ee
One might think that this indicates that the fuzzball is really small in some sense; it has just the natural minimum radius set by string theory. But such is not the case. In the NS1-P system that we are looking at $e^\phi$ becomes very small at the fuzzball surface. Thus the string tension becomes very low in planck units; in other words, the string becomes very long and `floppy'. Thus we should interpret (\ref{tseven}) as telling us that string length is very large, not that ${\cal R}$ is very small.  

This may sound more a matter of language rather than physics, but we can make it more precise by looking at the NS1-NS5 system which is obtained from NS1-P by dualities. It is a general result that the area of a surface $r=const$ (measured in planck units) does not change upon dualities. To see this, note that the Einstein action in D spacetime dimensions scales with the metric as follows
\be
S\sim {1\over G_D}\int d^D x \sqrt{-g}R \sim {[g_{ab}]^{D-2\over 2}\over G_D}
\label{ssone}
\ee
This action must remain unchanged under S,T dualities. The hypersurface at fixed $r$ is $D-2$ dimensional. Under dualities the $D-2$ dimensional area scales as $[g_{ab}]^{D-2\over 2}$. We have $G_D=(l_p^{(D)})^{D-2}$. From the invariance  of (\ref{ssone}) we see that  under dualities the scalings are such that the area of the $D-2$ dimensional fuzzball boundary measured in D-dimensional planck units remains invariant.  This fact remains true whether we use a 10-D description or the dimensionally reduced 5-D one. 

 For the fuzzball boundary in the NS1-NS5 system we get
\be
{A_{10}\over 4 G_{10}}={A_5\over 4 G_5}\sim \sqrt{n_1n_5}
\ee
(We have re-labeled the charges as $n_1\r n_5, n_p\r n_1$  to give them their appropriate names in the NS1-NS5 geometry.)
Thus 
\be
{\cal R}_{\rm NS1-NS5}\sim (n_1n_5)^{1\over 6} l_p^{(5)}
\ee
But this time the dilaton does not become strongly negative near the fuzzball boundary; rather it remains order unity
\be
e^{-2\phi}\approx {Q_1\over Q_5}\sim {n_1g^2\alpha'^2\over n_5V}
\ee
To find ${\cal R}$ in terms of string length and other moduli we write the 10-D area-entropy relation
\be
{A_{10}\over 4G_{10}}\sim {{\cal R}_{\rm NS1-NS5}^3 V R\over g^2\alpha'^4} \sim \sqrt{n_1n_5}
\ee
to get
\be
{\cal R}_{\rm NS1-NS5}\sim [{g^2\alpha'^4\over VR}]^{1\over 3} (n_1n_5)^{1\over 6}
\ee
We thus see that $l_p$ and $l_s$ are of the same order (their ratio does not depend on $n_1, n_5$) while ${\cal R}$ grows much larger than these lengths as the number of quanta in the bound state is increased. 

Thus we see that comparing the bound state size to the string length is not a duality invariant notion, while comparing it to the 5-D planck length is. In planck units the bound state size grows with charges. In string units, it also grows with charges in the NS1-NS5 duality frame, while it remains $l_s$ in the NS1-P duality frame. The latter fact can be traced to the very small value of $e^\phi$ at the fuzzball boundary, which makes the local string length very large in this case.

Sen \cite{sen} looked at the naive geometry of NS1-P, and noticed that the curvature of the string metric became string scale at $r\sim\sqrt{\alpha'}$, the same location that we have found in (\ref{ffone}) as the boundary of the fuzzball. He then argued that one should place a `stretched horizon' at this location, and then observed that the area of this horizon gave the Bekenstein relation (\ref{fftwo}). But if we look at the NS1-NS5 system that we obtain  by dualities, then the curvature remains {\it bounded and small} as $r\r 0$. The naive geometry is locally $AdS_3\times S^3\times T^4$ for small $r$, and  the curvature radii for the $AdS_3$ and the $S^3$ are $(Q'_1Q'_5)^{1/4}\gg\sqrt{\alpha'}$. So it does not appear that the criterion used by Sen can be applied in general to locate a `stretched horizon'. What we have seen on the other hand is that the naive geometry is not an accurate description for $r$ smaller than a certain value; the interior of this region is different for different states, and the boundary of this region satisfies a Bekenstein type relation (\ref{fftwo}). Further, we get the same relation (\ref{fftwo}) in all duality frames.

We have considered the compactification $T^4\times S^1$, but we could also have considered $K3\times S^1$. Suppose we further compactify a circle $\tilde S^1$, thus getting a 2-charge geometry in 3+1 noncompact dimensions. In this case if we include higher derivative corrections in the action then the naive  geometry develops a horizon, and the area of this horizon correctly reproduces the microscopic entropy \cite{dab}. Order of magnitude estimates suggest that a similar effect will happen for the 4+1 dimensional case that we have been working with. 

 What are the {\it actual} geometries for say NS1-NS5 on $K3\times S^1$?  Recall that in the NS1-P system that we started with the NS1 string had 8 possible directions for transverse vibration. We have considered the vibrations in the noncompact directions $x_i$; a similar analysis can be carried out for those in the compact directions $z_a$ \cite{lmm}. But after dualizing to NS1-NS5 we note that the the solutions generated by the $x_i$ vibrations are constant on the compact $T^4$, and we can replace the $T^4$ by $K3$ while leaving the rest of the solution unchanged. (The vibrations along compact directions will be different in the $T^4$ and $K3$ cases, but since we are only looking for estimate of the fuzzball size we ignore these states in the present discussion.) In \cite{gm2} it was shown that the higher derivative terms do not affect the `capped' nature of the `actual' geometries. Thus the K3 case is interesting in that it provides a microcosm of the entire black hole problem: There is a naive geometry that has a horizon, and we have `actual' geometries that have no horizons but that differ from each other inside a region of the size of the horizon.  It would be interesting to understand the effect of higher derivative terms on the naive $T^4$ geometry.
 
 \subsubsection{3-charge states}
 
 The 2-charge hole has two length scales. The radius of the horizon scales as $ (n_1n_2)^{1/6}l_p\sim n_i^{1/ 3} l_p$ where
$n_1, n_2$ are the charges in any duality frame and $l_p$ is the 5-d planck length in that frame (this scaling gives the Bekenstein entropy from the horizon area). On the other hand the distortion of the metric due to charges reaches to a distance
$\sim Q_1^{1/ 2}, Q_2^{1/ 2}$ which scale as $\sim n_i^{1/ 2}$. Thus for large $n_i$ (the classical limit) the horizon radius is much larger than the planck length (which suggests that we have the physics of a black hole) but the  `charge radius' is even larger. For the 3-charge hole the radius of the horizon and the `charge radius' both scale as
$n_i^{1\over 2}$, so we can take a classical limit where both scales are visible in the classical geometry. It would be satisfying to see the same `fuzzball' ideas emerge for 3-charge microstates, since we would then be describing the kind of black hole that has been long studied by general relativists. We do not know how to make all microstates of the 3-charge hole, but some selected families have been constructed \cite{3charge}. All these geometries agree with the `fuzzball' notion since they have no horizons, but instead are `capped' at small $r$. The constructed states are however not generic in that they have a large amount of rotation. Several other constructions suggest that similar results will hold for generic states, in particular the work on supertubes and black rings which shows that black objects are not as unique as initially believed, and which suggests ways to construct three charge states \cite{3chargeother,3chargenew,bk1,benawarner,gimonh}.

\section{`Fractionation' and the size of bound states}
\label{frac}\setcounter{equation}{0}

We have seen above that the 2-charge bound state has a size that grows with the charges; further, the area of the boundary and the degeneracy of the state satisfy a Bekenstein type relation $A/4G\sim S$.  What causes the bound state to `swell up' in this fashion, and become much bigger than a fixed length like planck length or string length? For 2-charge extremal states we constructed the states explicitly, and traced the size to the fact that momentum charge could only be carried on an NS1 by transverse vibrations of the NS1. But we would like to have a more general understanding of the underlying physics, so that we can extrapolate the ideas here to more general black holes.

In this section we will use somewhat heuristic arguments to arrive at a picture of bound states in string theory. 
The discussion of this section is based on \cite{emission}. The key idea will be `fractionation', which we encountered already when studying momentum waves on a string. If we have a graviton on a circle of radius $R$ then its energy and momentum  will be $n/R$ with $n$ an integer.  If on the other hand we have a string wound $m$ times on this circle then binding the graviton to the string gives traveling waves, which have energy and momenta occurring in units $n/(mR)$. This `fractionation' of the momentum leads to a large degeneracy of possibilities when we distribute the momentum among the harmonics, so it leads to the large entropy of the system. Does it also help us understand the large size?

At first it may appear that this effect is a rather simple and mundane one; it would happen for any string in any theory, so it does not seem to have much to do with string theory or black holes. But in string theory we have dualities that map fractional momentum modes to `fractional branes', which will have fractional tension  $T/m$. Thus for large $m$ they can be very `stretchable, floppy objects', and can thus extend far to give the bound state a large `size'. Since the tension goes down with the number of quanta in the bound state, we may hope that the size of the state might  always keep up with the horizon radius, however big a black hole we may try to make. In that case we will not have the usual black holes, but `fuzzballs'.

We will now use the computations done in earlier sections to offer some justification for the above picture of bound states.
We will  argue that fractional branes must exist, that they can be expected to stretch out to macroscopic distances, and that this distance will be of order the horizon radius for the 3-charge extremal black hole that we had constructed.

\subsection{Exciting the NS1-P extremal state}\label{exci}

We have seen that the three charges NS5,NS1,P can be permuted by dualities. Consider a 2-charge system, and let these charges be NS1,P. The extremal state is given by a NS1 wound $n_1$ times around the $S^1$ parametrized by $y$, with $n_p$ units of momentum running along the positive $y$ direction. The entropy of this extremal state is
\be
S_{ex}=2\pi\sqrt{2}\sqrt{n_1n_p}
\ee
Now consider adding some energy $\Delta E$ to this system so that it becomes slightly non-extremal. Where does this extra energy go? One might think that it goes towards creating extra vibrations of the NS1. Since we add no net momentum we must have
\be
n_p\r n_p+{R \Delta E\over 2}, ~~~\bar n_p={R\Delta E\over 2}
\ee
which implies an entropy
\be
S_{NS1-P+P\bar P}=2\pi\sqrt{2}[\sqrt{n_1(n_p+{R \Delta E\over 2})}+\sqrt{n_1{R \Delta E\over 2}}]
\ee
The subscript on the entropy tells us that the system is NS1-P and that it has been assumed that the additional energy has gone to creating $P\bar P$ pairs:
\be
{\rm NS1~P}~+~\Delta E~~\r ~~ {\rm NS1~P}~+~ (P~\bar P)
\label{pp}
\ee

Let us also write down the expected emission rate from this near-extremal system. The $P\bar P$ vibrations can collide and leave the string as massless bulk quanta. Let us look at the bulk mode we considered before -- the component $h_{12}$ of the metric, where $z^1, z^2$ are directions in the $T^4$. In section(\ref{twol2}) we had computed the emission from vibrations on the effective string of the NS1-NS5 bound state, but the computation applies equally well to the NS1 string that we have here. The emission rate is given by \cite{emparan}
 \bea
 \Gamma_{NS1-P+P\bar P}&=&\int {d^4 k\over (2\pi)^4}~(2\pi\omega G_5 L_T)~\rho_R({\omega\over 2})\rho_L({\omega\over 2}) \nn
 &=&\int {d^4 k\over (2\pi)^4}~(4\pi^2\omega G_5  Rn_1)~{1\over (e^{\omega\over 2 T_R}-1)(e^{\omega\over 2 T_L}-1)}
 \label{em1}
\eea
where we have set the length of the string to be $L_T=2\pi R n_1$. The temperatures $T_R, T_L$ are given by (\ref{temps}) but now with $f_B=f_F=8$  since there are 8 possible transverse vibrations of the string
\be
T_R=\sqrt{{({n_p\over R}+{\Delta E\over 2})\over 2\pi^2 R n_1}}, ~~~T_L=\sqrt{{({\Delta E\over 2})\over 2\pi^2 R n_1}}
\ee

\subsubsection{A puzzle}

Are these the correct near extremal entropy and emission rate? The emission process we have considered is
\be
{\rm NS1+P+nonextremality} ~~\r ~~ h_{12}
\ee
By S,T dualities we can map the NS1-P to NS5-NS1, while $h_{12}$ remains $h_{12}$, which gives the process
\be
{\rm NS5+NS1+nonextremality} ~~\r ~~ h_{12}
\ee
But this is a case that we have studied before, in section(\ref{twol2}). In that calculation we had a near-extremal NS1-NS5 system, and the energy above extremality went to creating $P\bar P$ pairs. The emission rate from (\ref{gmicro}) is
\bea
 \Gamma_{NS5-NS1+P\bar P}
 &=&\int {d^4 k\over (2\pi)^4}~(4\pi^2\omega G_5  R'n'_1n'_5)~{1\over (e^{\omega\over 2 T'_R}-1)(e^{\omega\over 2 T'_L}-1)}
 \label{em2}
\eea
where now the length of the effective string is $L_T=2\pi R' n'_1 n'_5$. The temperatures  $T'_R, T'_L$ are given by (\ref{temps})
 \be
T'_R=\sqrt{{({\Delta E\over 2})\over \pi^2 R' n'_1n'_5}}, ~~~T'_L=\sqrt{{({\Delta E\over 2})\over \pi^2 R' n'_1n'_5}}
\ee
We have denoted the radius of the $S^1$ by $R'$ since it will be related by dualities to the initial radius $R$. The charges are
also related by dualities
\be
n'_1=n_p, ~~~n'_5=n_1
\ee
But the radiation rate $\Gamma$ should be invariant under the dualities. This, however, is manifestly {\it not } so, if (\ref{em1}) gives the NS1-P result and (\ref{em2}) gives the NS1-NS5 result. In (\ref{em2}) we have equal $T_R, T_L$. In (\ref{em1}) we have {\it unequal} $T_R, T_L$, with the ratio diverging as we go closer to extremality $\Delta E\r 0$. Thus the functional forms of (\ref{em1}) and (\ref{em2}) are not the same.

In fact there is another more basic difference between the two emission rates. Recall that in the microscopic picture of the NS1-NS5 system the effective string can vibrate only in the plane of the NS5, so at leading order we can absorb a graviton like $h_{12}$ (which is a scalar in 4+1 D) but not a graviton like $h_{\mu\nu}$ where $\mu, \nu$ are two of the {\it noncompact} directions. $h_{\mu\nu}$ {\it is} absorbed at higher order in $\omega$, by exciting fermions in addition to the bosons on the effective string; these fermions carry spin under the rotation group of the noncompact directions.  This fact accords well with the gravity picture, where the fact that $h_{\mu\nu}$ is a spin 2 particle in 4+1 D implies that it feels an angular momentum barrier and its cross section is suppressed by powers of $\omega$. Thus in both microscopic and gravity pictures, at leading order in $\omega$ only the $h_{ij}$ with $i,j$ in the $T^4$ are absorbed.

But in the NS1-P case all 8 transverse directions of the string are on the same footing, and thus in  the microscopic  computation we get the same cross section for all components $h_{MN}$, where $M,N$ run over the 8 spatial directions transverse to the $S^1$. This does {\it not} accord with the gravity computation for a 2-charge system.

So something has gone wrong, (\ref{em1}) and (\ref{em2}) are {\it not} computations of emission in duality related processes.

\subsubsection{Resolving the puzzle}
 
The reason for the mismatch is not hard to find.  In the NS1-NS5  system that we had studied before the energy above extremality went to creating $P\bar P$ excitations
\be
{\rm NS1~NS5}~+~\Delta E~~\r ~~{\rm NS1~NS5} ~+~(P~\bar P)
\label{ns1ns5}
\ee
Permuting charges by duality we get the process
\be
{\rm NS1~P}~+~\Delta E~~\r ~~{\rm NS1~P} ~+~(NS5~\overline{NS5})
\label{ns1p}
\ee
which is {\it not} (\ref{pp}).

The model with excitations (\ref{ns1ns5}) gave us correctly the near extremal entropy and the correct Hawking radiation, so it is a model that we trust. We are then forced to accept the process (\ref{ns1p}) by duality. It may be hard to visualize how pairs of $NS5~\overline {NS5}$ can be created, or how they can interact to give rise to the emitted radiation, but since we have arrived at (\ref{ns1p}) by duality we will investigate the energy scales involved and see what we can learn.  Adopting (\ref{ns1p}) certainly removes the puzzle we faced above. The number of $NS5$ and $\overline {NS5}$ will be equal in (\ref{ns1p}) (there is no net NS5 charge), so the left and right temperatures will be equal, as was the case in the near extremal NS1-NS5 case. We do not have a simple model like  (\ref{inter}) to give the decay of $NS5~\overline{NS5}$ pairs to gravitons, but duality assures us that the correct spins will be emitted, and we do not have the obvious contradiction that we faced with the excitations (\ref{pp}) where all spins were emitted equally. 

What then is the role of (\ref{pp})? It certainly looks a correct description for excitations of the weakly coupled string. The mass of the lightest possible $P\bar P$ pair is
\be
\Delta E^{NS1P}_{P\bar P}~=2(E_p){1\over n_1}~=~2({1\over R}){1\over n_1}
\label{expp}
\ee
where $E_p=1/R$ is the energy of a momentum mode, the $2$ arises because we must create these modes in pairs so as to add no net P charge, and the factor ${1\over n_1}$ is the `fractionation', which will be crucially important since we are interested in the limit where all charges will be large.  Since we have one power of the charge in the denominator we will call this a case of `single fractionation'. 

In the process  (\ref{ns1ns5}) the energy of the lightest  excitation pair is
\be
\Delta E^{NS1NS5}_{P\bar P}~=~2(E_p){1\over n'_1n'_5}~=~2({1\over R'}){1\over n'_1n'_5}
\label{fullfrac}
\ee
Since there are two charges in the denominator on the RHS we call this a case of `double fractionation. 

Since (\ref{ns1p}) is dual to (\ref{ns1ns5}) we will have  `double fractionation' in (\ref{ns1p}) as well
\be
\Delta E^{NS1P}_{NS5\overline{NS5}}~=~2(m_5){1\over n_1n_p}~=~2({VR \over g^2\alpha'^3}){1\over n_1n_p}
\label{exns}
\ee
where $m_5$ is the mass of a NS5 brane. 

Which excitation will be preferred, (\ref{expp}) or (\ref{exns})? We may expect that the lighter excitation will generate more entropy for the same total energy and so will be the preferred excitation; we will investigate the issue of entropy in more detail below.  (\ref{exns}) has a factor $1/g^2$, so (\ref{expp}) is the lighter excitation for $g\r 0$, which makes sense because this is the free string limit and the excitations should be just vibrations of the string. But (\ref{exns}) has `double fractionation' while (\ref{expp}) has only single fractionation, so for given $g$ and sufficiently large charges (\ref{exns}) will be the lighter excitation.

We will now see that {\it whenever we are in the `strong gravity' domain (where we make a black hole) then (\ref{exns}) will be the lighter excitation}. 

We are looking at the NS1-P system. The NS1 string has to carry $n_p$ units of momentum. If the momentum is in low harmonics then the amplitude of vibration will be large and the string will spread over a large region; if the momentum is in high harmonics then the string will have a small transverse spread. For the generic vibration mode the transverse spread
$\Delta x$
was found in (\ref{ffone}), and we will use this value for the transverse size of the string state. On the other hand the gravitational field of the string is described by the quantities $Q_1, Q_p$ which occur in the metric as $\approx {Q_1\over r^2}, {Q_p\over r^2}$. We will say that the string strongly feels its own gravity if
\be
(\Delta x)^2 ~ \lesssim~Q_1, ~Q_p
\label{strong}
\ee
so that the size of the string is smaller than the reach of the gravitational effects of both kinds of charges. Since $Q_1, Q_p$
may be unequal, we   want $\Delta x$ to be smaller than both these length scales, a requirement that we can  encode by writing
\be
(\Delta x)^2 ~\lesssim~{Q_1 Q_p\over Q_1+Q_p} 
\ee
Using (\ref{ffone}), the values (\ref{sixt}) for $Q_1, Q_p$ and noting that the mass $M$  of the string is 
\be
M={n_1 R\over \alpha'}+{n_p\over R}
\ee
we find that the condition that the string feels its own gravity (rather than be a free string in flat space) is
\be
\alpha'~\lesssim~{g^2\alpha'^3\over VR} {n_1n_p\over M}
\label{condition}
\ee
Now consider the ratio of the two kinds of excitations  that we wished to compare. In (\ref{expp}) we had excited only the momentum modes, but to be more precise we note that by T-duality the winding modes can be excited as well. The excitation levels of the string are in fact given by the relation (\ref{massformula}) with $T={1\over 2\pi\alpha'}$. This gives
\be
2M\delta M~=~{4\over \alpha'}\delta N_L~=~{4\over \alpha'}\delta N_R
\ee
For the lowest excitation we set $\delta N_L=\delta N_R=1$ and find
\be
\Delta E^{NS1P}_{\rm vibrations}~=~\delta M~\sim ~ {1\over \alpha' M}
\ee
We then find
\be
 {\Delta E^{NS1P}_{NS5\overline{NS5}}\over \Delta E^{NS1P}_{\rm vibrations}  }~\sim~{RVM\over g^2\alpha'^2 n_1n_p} 
\ee
Comparing with (\ref{condition}) we find that whenever we are in the strong gravity regime (\ref{strong}) we have
\be
  {\Delta E^{NS1P}_{NS5\overline{NS5}}\over \Delta E^{NS1P}_{\rm vibrations}  } ~\lesssim 1
\ee
so that the fractional NS5 brane pairs are lighter than vibrations of the string.

\subsubsection{Entropy and phase transitions}

Even though we have found that the fractional NS5 brane pairs may be lighter, this does not mean that they are the preferred excitation; we have to check that exciting them creates more entropy for the same energy. Entropy measures the log of the volume of phase space. So if we require
\be
{\Delta S_{NS5\overline{NS5}}\over \Delta S_{vibrations}  }~>~1
\ee
then  (after the system comes to equilibrium) we will find that the energy of excitation is carried by the fractional 5-brane pairs in preference to vibrations of the string.

To understand the role of entropy in this story   consider first  the NS1-NS5 bound state discussed in section(\ref{extr2}); the present system NS1-P is of course related to this by duality. In the NS1-NS5 system we had an effective string with total winding number $n_1n_5$. We could partition this effective string into `component strings', where the component strings have winding numbers satisfying (\ref{six}). The different ways to make these partitions, together with the different possible fermion zero modes on the component strings, gives the 2-charge entropy (\ref{sixfollow}). 

But the generic component string in a generic 2-charge state has winding number $\sim \sqrt{n_1n_5}$ (we can see this by applying duality to (\ref{thir})), so it does {\it not} give the fractionation (\ref{fullfrac}) for $P\bar P$ excitations. It is true that if we want to get the maximal possible entropy from the 
$P\bar P$ excitations then we should have the maximal possible fractionation (\ref{fullfrac}); this partitions each momentum mode into $n_1n_5$ pieces and the count of these partitions gives the entropy. But to get this maximal fractionation we have to join all the component strings into one single long component string with winding number $n_1n_5$, so we {\it lose} the `2-charge entropy' that came from the different possible partitions of the winding number. If there is a very small excitation energy, giving very few $P\bar P$  pairs, then we would suffer a loss of entropy if we make the single long string; on the other hand if there was a sufficient amount of $P\bar P$ excitation then it will be {\it more} profitable to lose the `2-charge entropy' and make one single long component string, then {\it gain} the entropy coming from the maximally fractionated $P\bar P$  pairs. Since these pairs are the `third charge' in this problem, we call this latter entropy  `3-charge entropy'. To summarize, once we cross a certain threshold of excitation energy the system prefers to rearrange its degrees of freedom so that the entropy is `3-charge' rather then `2-charge': The entropy obtained from different ways of partitioning the NS1-NS5 effective string is `sacrificed', and the single long string which results then  generates a higher entropy by maximally partitioning the {\it third} charge P.

Let us now return to our NS1-P system, and apply the above principles. All we need do is permute under duality the charges involved in the above story. Let the excitation energy be $\Delta E$. If we put this into `2-charge' degrees of freedom then we have the entropy of a fundamental string
\bea
S_{vibrations}&=&2\pi\sqrt{2}\sqrt{N_R+\delta N_R}+2\pi\sqrt{2}\sqrt{\delta N_L}\approx 2\pi\sqrt{2}\sqrt{N_R}+2\pi\sqrt{2}\sqrt{\delta N_L}\nn
&=&2\pi\sqrt{2}\sqrt{n_1n_p}+2\pi\sqrt{2}\sqrt{{M\alpha'\over 2}\Delta E}=S_{ex}+2\pi\sqrt{2}\sqrt{{M\alpha'\over 2}\Delta E}
\label{en1}
\eea
where $S_{ex}$ is the extremal 2-charge entropy. Here the extra energy $\Delta E$ has gone to just exciting the two charges NS1-P that we started with, so this is `2-charge entropy'. 

The `3-charge entropy' will arise if we excite the third charge in the story, the fractional NS5 pairs. We assume, using duality,
that we sacrifice the `2-charge entropy', getting the NS1-P charges into a specific state which then manages to fractionate the NS5 charges by the maximal amount (\ref{fullfrac}). This gives
\be
S_{NS5\overline{NS5}}=2\pi\sqrt{n_1n_pn_5}+2\pi\sqrt{n_1n_p\bar n_5}=4\pi\sqrt{n_1n_p( {g^2\alpha'^3\over VR})({\Delta E\over 2})}
\label{en2}
\ee
The entropy (\ref{en2}) will become comparable to (\ref{en1}) (in which the dominant contribution is $S_{ex}$) when
\be
{g^2\alpha'^3\over VR} \Delta E\sim 1
\label{compare}
\ee
Since the nonextremal energy shows up in the geometry (\ref{fullmetric}) through the parameter $r_0$, it is worth asking what value of $r_0$ the $\Delta E$ in (\ref{compare}) corresponds to. Using (\ref{energy}) with $\cosh 2\alpha\approx\sinh2\alpha, \cosh 2\sigma\approx \sinh 2\sigma, \gamma=0$ we find
\be
\Delta E=E-E_{extremal}={RV\over 2g^2\alpha'^4}r_0^2
\ee
Using (\ref{compare}) we find 
\be
r_0\sim\sqrt{\alpha'}
\ee
But this is just the `horizon' radius of the extremal 2-charge `fuzzball' (\ref{ffone}).

\subsubsection{Summary}\label{summ}

Let us summarize the discussion of this subsection. Start with extremal NS1-P and add a small amount of energy. At very weak coupling  we just get additional vibrations of the string. But if $g$ is large enough that the string strongly feels its own gravity, then there is an excitation that is lighter than the lowest vibration mode of the string:  A `maximally fractionated' $NS5\overline {NS5}$ pair. But to make such a maximally fractionated pair possible the NS1-P string needs to be in a specific state, so we would need to lose `2-charge entropy'. Suppose we make the added energy $\Delta E$ large enough so that the horizon of the geometry (\ref{fullmetric}) becomes larger than the radius of the 2-charge extremal NS1-P `fuzzball'. Then `maximal fractionation' becomes entropically favored, and the excitation energy is carried by  
$NS5\overline {NS5}$  pairs. Our computations of near extremal entropy and radiation (sections (\ref{them}) and (\ref{abso})) were compared with the properties of the metric (\ref{fullmetric}). In this metric the classical limit of large charges $n_i\r \infty$ has been taken. The 2-charge fuzzball radius $\Delta x$  grows as $n_i^{1/3}$. In (\ref{fullmetric}) we have the `charge radii' $\sqrt{Q_i}$ and the nonextremality parameter $r_0$, and the near extremal limit corresponds to $r_0\ll \sqrt{Q_i}$. But the scales $\sqrt{Q_i}$ grow as $n_i^{1/2}$. Even though we may choose $r_0/\sqrt{Q_i}\ll1$ for our near-extremal calculations, we will still have $r_0\gg \Delta x$ in the classical limit.

 {\it Thus to understand the dynamics of a black hole like (\ref{fullmetric}) with NS1-P charges we need to think of fractional $NS5\overline {NS5} $  pairs.\footnote{The effect of strong self-gravity on a string was also discussed in \cite{hp} from a slightly different point of view; it was noted that the entropy of a string state agreed with the entropy of a black hole at the `correspondence point' where the horizon radius just equalled the string scale.
 We have argued here that such correspondence points are points of phase transition where the microscopic degrees of freedom completely change character.}}

\subsection{Squeezing a black hole}\label{sque}
 
Consider the 3-charge NS1-NS5-P black hole, with  small amount of nonextremality. We assume for convenience that $Q_p$ is smaller than $Q_1, Q_5$ so the extra energy goes mainly towards creating $P\bar P$ pairs
\be
{\rm NS1~NS5~P}~+~\Delta E~~\r ~~{\rm NS1~NS5~P} ~+~(P~\bar P)
\label{ns1ns5ppp}
\ee
The entropy is
\be
S^{NS1NS5P}_{P\bar P}~=~2\pi\sqrt{n_1n_5(n_p+\Delta n_p)}+2\pi\sqrt{n_1n_5\bar n_p}\equiv S_{ex}+\Delta S
\label{en21}
\ee
where the extra entropy $\Delta S$ is much smaller then the extremal entropy $S_{ex}=2\pi\sqrt{n_1n_5n_p} $ since we are assuming that we are close to extremality.

Now imagine that one of the noncompact spatial directions (say $x^4$) is compactified, on a circle of radius $\tilde R$. If this circle is large
(i.e. $\tilde R$ is much greater than all other length scales in the problem) then we have the near extremal 3-charge hole as above. But now imagine reducing $\tilde R$. If $\tilde R$ is small then we should be thinking of the {\it four} charge black hole in 3+1 noncompact directions. In section(\ref{the4}) we had seen that the four charge hole has charges NS1, NS5, P, KK where KK stands for Kaluza-Klein monopoles with $x^4$ as the nontrivially fibered circle. If we have just the charges  NS1-NS5-KK and a small amount of nonextremality then we excite (fractional) $P\bar P$ pairs (eq. (\ref{4chargeen}))
\be
{\rm NS1~NS5~KK}~+~\Delta E~~\r ~~{\rm NS1~NS5~KK} ~+~(P~\bar P)
\label{ns1ns5kk}
\ee
By duality we can permute the four charges, so if we start with NS1-NS5-P charges and add a little nonextremality then we should excite (fractional) $KK\overline {KK}$ pairs
\be
{\rm NS1~NS5~P}~+~\Delta E~~\r ~~{\rm NS1~NS5~P} ~+~(KK~\overline {KK})
\label{ns1ns5p}
\ee
The mass of a KK monopole is
\be
m_{KK}={RV\tilde R^2\over g^2\alpha'^4}
\label{mkk}
\ee
so the entropy of this near extremal system will be
\be
S^{NS1NS5P}_{KK\bar KK}~=~2\pi\sqrt{n_1n_5n_pn_{KK}}+2\pi\sqrt{n_1n_5n_p\bar n_{KK}}
=4\pi\sqrt{n_1n_5n_p ({\Delta E\over 2 m_{KK}})}
\label{en22}
\ee
Which is the correct description of microscopic excitations, (\ref{ns1ns5ppp}) or (\ref{ns1ns5p})? 
This, we expect, depends on which entropy is higher, (\ref{en21}) or (\ref{en22}). Noting that (\ref{en21}) is dominated by the contribution $S_{ex}$ we find that these entropies become comparable when \cite{emission}
\be
\Delta E~\sim~ m_{KK}~\sim~ {RV\tilde R^2\over g^2\alpha'^4}
\label{energy2}
\ee
For larger energies above extremality, or equivalently, when $\tilde R$ is smaller than the value given by the above relation, we will have the excitations (\ref{ns1ns5p}) in preference to the excitations (\ref{ns1ns5ppp}).
So again we see a `phase transition' where the microscopic degrees of freedom change character when a parameter ($\tilde R$) is moved past a critical point. 

It is interesting to ask what values of the nonextremality parameter $r_0$ the critical energy (\ref{energy2}) corresponds to.
In the metric of the 3+1 dimensional hole (\ref{4chargemetric}) the energy above extremality is
given by (using (\ref{energy4}))
\be
\Delta E\approx {RV\tilde R r_0\over g^2\alpha'^4}
\ee
Equating this to (\ref{energy2}) we find
\be
r_0\sim \tilde R
\label{length1}
\ee
Note that $\t R$ need not be small (i.e. it can be much bigger than planck length or string length) so we see that
our fractional monopole pairs can stretch out to macroscopic distances.

In the 4+1 dimensional metric all charges are nonzero so it takes a little more effort to compute the energy above extremality corresponding to (\ref{energy2}) (we hold the $\hat n_i$ fixed and change $r_0$). We find 
\be
\Delta E\approx {RVr_0^4\over 8g^2\alpha'^4}\, ( Q_1^{-1}+Q_5^{-1}+Q_p^{-1})
\ee
Equating this to (\ref{energy2}) we find
\be
r_0^4\sim {\tilde R^2\over  Q_1^{-1}+Q_5^{-1}+Q_p^{-1}}
\label{length2}
\ee
The physics here seems similar to that in the Gregory-Laflamme transition \cite{gl}. In the Gregory-Laflamme transition we find that as we increase the size of a circle in the spacetime we go from a black brane wrapping this circle to a black hole that does not  fill the compact circle.  In our microscopic computation above we found that if we increase the size of the circle $\tilde S^1$ then we go from an object which has KK monopole-pairs wrapping the circle to an object that does not make use of this circle in distributing its energy. It would be interesting to see if the length scales (\ref{length1}),(\ref{length2}) can be understood as a case of the Gregory-Laflamme type \cite{gl}, by analyzing the microscopic entropy in terms of the energy and `pressures' created by the brane bound state \cite{obers}.

\subsection{Estimating the size of the NS1-NS5-P bound state}\label{esti}

Let us put together all that we have learnt to make an ambitious attempt: We will try to obtain a crude estimate of the `size' of the 3-charge extremal bound state. We have already constructed explicitly the 2-charge states, and found that their radius $R_s$ is not `small', i.e. $R_s$ is not a fixed scale like planck length or string length, but a length that grows with the charges in the bound state in such a way that the surface area of a sphere at $r\sim R_s$ gives the Bekenstein entropy of the 2-charge state. We do not know how to make the generic 3-charge state, but we will try to use the concepts of fractionation and the origin of black hole entropy to get some idea of the dynamics that could govern the `size' of the 3-charge state.

We proceed in the following steps:

\medskip

(a) Start with the compactification $M_{9,1}\r M_{4,1}\times T^4\times S^1$, and construct the NS1-NS5-P extremal bound state as before. Let this bound state be placed with its center at $r=0$. Bring a test quantum to a location $|\vec r|=R_d$ near the bound state. At what value of $R_d$ will the test quantum directly `feel' the bound state? If $R_d$ is very large then the test quantum will only feel the usual long range fields like the gravitational field produced by the bound state. But as we reduce $R_d$ there might be a critical value at which brane-antibrane pairs `stretch out and touch' the test quantum. If such a phenomenon were to occur, we would  say that $R_d$ is a measure of the physical size of the 3-charge bound state.

\medskip

(b) If brane-antibrane pairs have to be created to stretch out and affect the test quantum then we need energy to create such pairs. Where can this energy come from? We have deliberately made things as hard for ourselves as possible, by taking an extremal  NS1-NS5-P bound state; all the energy of this state is accounted for by the charges it carries, so there is nothing extra that can be used to produce the pairs. But the fact that we bring the test quantum to within a distance $R_d$ means that we have an energy $\Delta E>{1/ R_d}$ above extremality in the combined system of bound state plus test quantum. (For the test quantum $E=\sqrt{m^2+p^2}>p\sim 1/R_d$.) We set 
\be
\Delta E={1\over R_d}
\label{en3}
\ee
Note that we are hoping to get a macroscopic value for the length $R_d$ at the end of the computation, so this energy is very small. 

\medskip

(c) We do not have a good picture of what it means for branes `to stretch out and touch' the test quantum. So let us modify the problem somewhat. We put the bound state in a periodic box of length $R_d$, and add the energy (\ref{en3}) to the system. If the resulting excitations do not feel the size of the box then we would say that the `size' of the bound state is much smaller than the box size $R_d$. On the other hand if the excitation generates branes that wrap around the compact direction of the box then we would say that the bound state size is {\it larger} than $R_d$.

At this stage  we have now brought the problem to a form that can be tackled by the tools that we have already developed. We have a compactification $M_{9,1}\r M_{3,1}\times T^4\times S^1\times \tilde S^1$, with $\tilde S^1$ having radius $R_d$. We have a state with three large charges NS1-NS5-P, and a small amount of nonextremality. From (\ref{ns1ns5p}) we see that we can use the nonextremal energy to create fractional $KK\overline{KK}$ pairs. These monopoles have the  direction $\tilde S^1$ as the nontrivially fibered direction, so the produced pairs `wrap' around the circle of size $R_d$ and are thus certainly sensitive to the `box size'. The mass of a monopole grows with $R_d$ as $R_d^2$, so one might think that such monopole pairs are hard to produce with the small energy (\ref{en3}), but  fractionation can make these pairs light, as we will soon see.

\medskip

(d) One last step before we start computing. We want not only that the fractional $KK\overline{KK}$ pairs {\it can} be generated, we want that it should be likely that they {\it are} generated. Concretely, we demand that using the energy (\ref{en3}) to generate the pairs should lead to an  entropy increase $\Delta S\ge 1$. Since entropy measures the volume of phase space, this would imply that
\be
{{\rm Volume~ of ~phase ~space~ with } ~KK\overline{KK}~{\rm pairs ~created}\over {\rm Volume~ of ~phase~ space ~{\it without}~ pairs}}~\ge ~e
\ee
 so it is more likely than not that we create the pairs. Thus we set on ourselves the requirement
\be
\Delta S=1
\ee

\medskip

(e) Since the key effect here will be fractionation, let us pause for a moment to consider the nature of fractionation that we expect. We have seen that the extremal 3-charge system has an entropy $S_{ex}=2\pi\sqrt{n_1n_5n_p}$.
This entropy comes because we can distribute the fractionated components of this bound state (fractional momentum modes in the description that we had used) in many ways, getting the entropy $S_{ex}$. On that other hand, if we {\it sacrifice} this entropy to make a special state of the NS1-NS5-P system, then we can get the new excitation -- KK monopoles -- to be `maximally fractionated' in units $1/n_1n_5n_p$. In the computation of section(\ref{sque}) the `phase transition' occurred when it was advantageous to give up the entropy $S_{ex}$ and to gain instead the entropy of fractional $KK\overline {KK}$ pairs.

In the present case we are making a small perturbation to the 3-charge NS1-NS5-P extremal state, so we do not expect that 
there will be a complete rearrangement of the 3-charge state. Rather, we expect that a small fraction $\mu$ of the excitations will `bind together' to make a special configuration, and the rest will exhibit the entropy of the 3-charge state. Specifically
\bea
S&=&S_{3-charge}~+~S_{4-charge}\nn
&=&[\,2\pi\sqrt{n_1n_5((1-\mu)n_p)}\,]+[\,2\pi\sqrt{n_1n_5 (\mu n_p)n_{KK}}+2\pi\sqrt{n_1n_5 (\mu n_p)\bar n_{KK}}\,]
\label{fullen}
\eea
where
\be
n_{KK}+\bar n_{KK}=2n_{KK}={\Delta E\over m_{KK}}
\label{nkk2}
\ee
When the system comes to equilibrium the parameter $\mu$ will get set to the value that gives the maximum entropy. Extremising (\ref{fullen}) with respect to $\mu$ we find that the optimal value of $\mu$ is given through
\be
{\mu\over 1-\mu}=4n_{KK}~\r~\mu\approx 4n_{KK}
\label{mukk}
\ee
where in writing the approximation we have noted that we expect $\mu<<1$ at the end of the computation.
(Note that $n_{KK}$ will be fractional, so it can be much less than unity.)  Substituting this value of $\mu$ in (\ref{fullen}) we find
\be
S\approx 2\pi\sqrt{n_1n_5n_p}+4\pi\sqrt{n_1n_5n_p}~n_{KK}\equiv S_{ex}+\Delta S
\ee
Setting $\Delta S=1$ gives
\be
\mu={1\over \pi\sqrt{n_1n_5n_p}}
\ee
Relating $\mu$ to $n_{KK}$ (eq. (\ref{mukk})), $n_{KK}$   to $\Delta E/2m_{KK}$ (eq.  (\ref{nkk2})), and then using (\ref{en3}),(\ref{mkk}) we find
\be
{1\over R_d}{g^2 \alpha'^4\over R_d^2 RV}\sim {1\over \sqrt{n_1n_5n_p}}
\ee
which gives \cite{emission}
\be
R_d\sim [{g^2\alpha'^3\sqrt{n_1n_5n_p}\over RV}]^{1\over 3}
\ee
as the length scale to which the fractional KK monopoles extend.
But the Schwarzschild radius of the 3-charge extremal hole is
\be
R_s =  [{g^2\alpha'^3\sqrt{n_1n_5n_p}\over RV}]^{1\over 3}
\ee
(This is easily verified by using $A/4G_5=S=2\pi\sqrt{n_1n_5n_p}$.) The agreement of the numerical
coefficient is just fortuitous, but it is interesting that even in our very crude estimate all the other parameters fall into their correct place and tell us that the size of the 3-charge extremal bound state is not small; i.e. not a fixed scale like planck or string length, but something that grows with the charges, and in fact is of order the horizon radius. This tells us that nonperturbative string theory effects can correct the geometry in the entire interior of the hole, and not just within planck distance of the singularity. 

\subsection{Summary}

The arguments of this section have been rough, unlike the concrete computations that we presented in earlier sections. But note that these arguments used little more than the results that we had found in our more rigorous calculations. In the process we have uncovered phenomenon like fractionation, phase transitions, low tension fractional branes and  large sized fuzzballs. Note that even for the 2-charge extremal NS1-P system we can attribute the nontrivial size to `fractionation'. Suppose we had a string with winding number $n_1$. If we put on this string a wave with wavenumber an integral multiple of $1/R$ then  all strands of the string move together, and there is no significant transverse size. But if we put the energy in a mode with fractional wavenumber, say $k=1/(n_1R)$, then we have seen that the strands separate and spread out over a large transverse region.

All this suggests that we must use radically new concepts to study bound states of large numbers of quanta in string theory. Thus string theory can offer a perspective on the information paradox that we could not have obtained by using our semiclassical notions of gravity. 

\section{Discussion}
\label{disc}\setcounter{equation}{0}

Let us summarize the ideas that we have developed. Gedanken experiments tell us that a black hole must have entropy $S_{bek}=A/4G$. In string theory we can make extremal black holes and count their states from the properties of strings/branes; the BPS nature of these systems enables this count to be made in a weak coupling domain and then continued without change to the coupling  where we expect  black holes. Interestingly, such computations work also for near extremal holes, presumably because the excitations are sufficiently `dilute' on the large system of branes that the interactions between excitations can be ignored to leading order. From the interaction of these excitations with the bulk modes we also found that the emission from near extremal bound states agreed with the low energy Hawking radiation expected from the corresponding holes.

All this did not resolve the information paradox; for that we need an understanding of the {\it structure} of black hole states, not just their count or dynamics deduced from a weak coupling domain. For 2-charge extremal states we found that the bound states were not small, but were `fuzzballs' of `horizon size'. We wrote the family (\ref{ttsix}) of classical geometries to describe these states, but in doing so we assumed that the string in the NS1-P picture was well described by a classical profile. In general we have  an energy eigenstate of the string in a vibration mode, and there will be quantum fluctuations about the mean. These fluctuations do not change the size of the region over which the string spreads; they just make the object `fuzzy' so we term it a `fuzzball'. For more details on quantum fluctuations and the classical limit see \cite{fuzz}. There are additional corrections which in the NS1-NS5 picture come from D-string winding modes in the $y$ direction; in \cite{gm2} it was shown that these produce bounded corrections and so do not change the `fuzzball' picture of the microstates.

We have argued that a basic idea is `fractionation': The property of  objects in string theory that when they make supersymmetric bound states they break up into smaller `bits'. It is this breakup that leads to the large entropy
of string states (which agrees with the Bekenstein entropy). We have suggested that this breakup also leads to the large `size' of bound states that gives `fuzzballs' instead of smooth geometries with horizons. We get low tension `fractional branes' which can easily stretch to large distances, and a rough analysis suggests that this distance will be of order the horizon length.

If such is the case then we change the picture of the black hole interior: Different microstates will be different states of the `fuzzball' and radiation leaving from the surface of the fuzzball can see the information encoded in the state just like the photons leaving a burning piece of coal see the state of the coal and so carry its information. This would therefore resolve the information paradox.

What is the significance of the `naive' geometry (like (\ref{one})) that we can write down as a solution of 
the classical field equations? It is tempting to think that this could be some kind of an `average' over the microstate
geometries. But as in any quantum mechanical system our fuzzball must be in any {\it one} of its possible states, so what can such an average mean?  One situation where {\it all} the states will be involved is where we put the system on a Euclidean time circle of length $\beta$; this gives thermal partition function in which all microstates run around the time circle.
This partition function may be represented by a Euclidean path integral. Suppose that this path integral is dominated by a saddle point configuration. Based on what we know about Euclidean black holes we expect that this will be the `cigar shaped' Euclidean solution which ends smoothly at the horizon and has no `interior'. 

We can now continue this geometry to the Lorentzian section, getting the geometry (\ref{one}). This geometry has a horizon, but it does not have any direct significance as a geometry in our approach; the actual Lorentzian geometries were the `fuzzballs' which had no horizons. One may however be able use the geometry (\ref{one}) to compute Green's functions that give ensemble averages over the black hole states \cite{shenker}. 

It is plausible that the ideas arising in black holes extend to other situations where we have a large number of particles interacting strongly, as for example in the early Universe. If we get long distance quantum effects in the cosmological setting then it may affect our understanding of the initial wavefunction of the Universe, the horizon problem and the way we compute the vacuum energy density.

As mentioned in the introduction, we have sought to review a few selected computations in string theory to bring out a certain perspective on black holes. The presented computations are a small fraction of the research results in the area, and we hope that they  will encourage the reader to look deeper at the field.  In particular recent progress  has been quite rapid.  There is a general way to construct all BPS 3-charge geometries, though this set includes bound states as well as unbound ones
\cite{benawarner}. For some specific families of bound states dual geometries have been constructed \cite{3charge}. Larger families of such geometries were found in \cite{3chargenew}. 3-charge supertubes offer a microscopic approach to the problem, and geometries for these have been suggested \cite{bk1}. The discovery of black rings has provided another new aspect of the problem of `hair' for black holes; one finds that black hole metrics are not as unique
as originally believed \cite{ring}. There is a way to add monopole charges to BPS states, which gives possible microstates for 4-dimensional systems \cite{monopole}. Work with generic members of the ensembles of microstates show that they might exhibit black hole like properties \cite{thermal2}. Some nonextremal state geometries have been constructed as well \cite{nonex1}. This progress suggests that there is a vast body of knowledge remaining to be uncovered in this area.

\section*{Acknowledgements}

I would like to thank Sumit Das, Stefano Giusto, Oleg Lunin, Ashish Saxena and Yogesh Srivastava who have been collaborators in various aspects of the work discussed here. I also thank Borun Chowdhury, Stefano Giusto and Yogesh Srivastava for helping to correct errors in the manuscript. This work is
supported in part by DOE grant DE-FG02-91ER-40690.

\end{document}